\documentclass[aps,pre,reprint,groupedaddress,showpacs,showkeys]{revtex4-1}

\usepackage{latexsym,amssymb,amsfonts,mathrsfs,xcolor,graphicx,amsmath}
\usepackage[american]{babel}
\usepackage[normalem]{ulem}

\newcommand{\az}[1]{\textcolor{black}{#1}}
\newcommand{\bh}[1]{\textcolor{black}{#1}}     
\newcommand{\mh}[1]{\textcolor{black}{#1}}

\begin{document}

\title{Microswimmers learning chemotaxis with genetic algorithms}
\author{Benedikt Hartl}
\author{Maximilian H\"ubl} 
\author{Gerhard Kahl}
\author{Andreas Z\"{o}ttl}
\email{andreas.zoettl@tuwien.ac.at}
\date{\today}

\affiliation{Institute for Theoretical Physics, TU Wien, Wiedner Hauptstr.\ 8-10, 1040 Wien, Austria}

\begin{abstract}
Various microorganisms and some mammalian cells are able to swim  in viscous fluids by performing nonreciprocal body deformations, such as rotating attached flagella or by distorting their entire body.
In order to perform chemotaxis, i.e.\ to move towards and to stay at high concentrations of nutrients, they adapt their swimming gaits in a nontrivial manner.
We propose a model  how microswimmers are able to autonomously adapt their shape in order to swim \az{in one dimension} towards high field concentrations using an internal decision making machinery modeled by an artificial neural network.
We present two methods to measure chemical gradients, spatial and temporal sensing, as known for swimming mammalian cells and bacteria, respectively.
\az{Using the NEAT genetic algorithm
surprisingly simple neural networks evolve
which control the
shape deformations of the microswimmer 
and allow them to navigate in static and complex time-dependent chemical environments.} 
By \az{introducing} noisy signal transmission in the neural network the well-known biased run-and-tumble motion
emerges.
Our work demonstrates that the evolution of a simple internal decision-making machinery, which we can fully interpret and is coupled to the environment, allows navigation in diverse chemical landscapes. These findings are of relevance for 
intracellular biochemical sensing mechanisms of single cells, or for the simple nervous system of small multicellular organisms such as \textit{C.~elegans}.

\end{abstract}
\maketitle

\section*{Introduction}

Microogranisms possess a huge variety of different self-propulsion strategies in order to actively swim through viscous fluids such as water, which is realized by performing periodic nonreciprocal deformations of their body shape \cite{Purcell1977,Lauga2009a,Elgeti2015,Zoettl2016}. 
In order to search for nutrients, oxygen, or light, they have developed  mechanisms to change their shape and hence their swimming direction abruptly.
An important example is the run-and-tumble motion of various bacteria such as \textit{Escherichia coli}  \cite{Berg1972,Lauga2016} or of the algae \textit{Chlamydomonas} \cite{Polin2009}. 
In order to perform chemotaxis, bacteria   use temporal information of chemical field concentrations  mediated by a time-dependent response function  which suppresses tumbling when swimming upwards chemical gradients \cite{Berg1972, 
Clark2005,Celani2010,Taktikos2013}. 
Some bacteria follow more diverse chemotactic strategies which can be related to their specific propulsion mechanisms \cite{Alirezaeizanjani2020}.
In contrast to bacteria, many  eukaryotic cells such as \textit{Dictyostelium}  \cite{Swaney2010,Levine2013}, leukocytes \cite{Artemenko2014} or cancer cells \cite{Roussos2011} are able to perform chemotaxis by adapting their migration direction  in accordance with the chemical gradient by spatial sensing with membrane receptors. 
From an evolutionary point of view, it remains elusive how  motility and chemotactic patterns evolved together, bearing in mind that both different prokaryotic and eukaryotic cells with  diverse self-propulsion mechanisms developed surprisingly similar chemotactic machinery \cite{Jarrell2008,Artemenko2014,Wan2020}.

In our work we  use machine learning (ML) techniques in order to investigate how  chemotaxis-based decision making can be learned and performed in a viscous environment. During past years
various ML approaches have become increasingly appealing in different fields \bh{of} physics, for example in material science, soft matter and fluid mechanics  \cite{Butler2018,Mehta2019,Brunton2020}. 
Unsupervised reinforcement learning (RL)
has been used in various biologically motivated active matter systems \cite{Cichos2020} to investigate optimum 
strategies, employed by smart, self-propelled agents: examples are to navigate in fluid flow \cite{Colabrese2017,Gustavsson2017,Alageshan2020,Qiu2020} and airflow \cite{Reddy2016}, in complex environments, external fields \cite{Palmer2017} and potentials \cite{Schneider2019}.
\az{Noteworthy,} two contributions have taken the viscous environment into account, namely one applying Q-learning to a three-bead-swimmer \cite{Tsang2018}, and one using deep learning to find energetically efficient collective swimming of fish \cite{Verma2018}.
Experimental realizations of  ML applied to self-propelled objects 
are navigation
of microswimmers \az{on}
a grid \cite{Muinos-Landin2018} or  macroscopic gliders learning to soar in the atmosphere \cite{Reddy2018}.

Here we address the problem, how a  microswimmer is able to
make decisions by adapting its shape in order to perform chemotaxis.
To employ adaptive swimming behavior, microswimmers need to be -- to a certain extent -- aware of both their environment and their internal physiological state.
Substituting the complex biochemical sensing machinery of unicellular organisms, or real sensory and motor neurons of small multicellular organisms such as \textit{C.~elegans}\az{,}
we therefore employ the evolution of a simple artificial neural network (ANN)\az{,} which is able to sense the environment and proposes actions to deform the body shape accordingly.
We introduce both spatial and temporal chemical gradient sensing leading to different decision making strategies and dynamics in chemical environments.

\section*{Results}
\subsection*{Microswimmer model}

As a simple model we use the so-called three-bead swimmer introduced originally by Najafi and Golestanian  \cite{Najafi2004}. It swims in a viscous fluid of viscosity $\eta$ via periodic, nonreciprocal deformations of two arms, connecting three aligned beads of radius $R$, located at positions $x_i$, $i=1,2,3$ (see top left panel in Fig.~\ref{Fig:1}). The central bead is connected to the outer beads  by two arms: their variable lengths $L_1$ and $L_2$ are extended and stretched by time-dependent forces $F_i(t)$ acting on the hydrodynamically interacting beads, which determine the bead velocities $v_i(t)$ \cite{Golestanian2008} (see SI Appendix). In this manner a force-free microswimmer (i.e., $\sum_i F_i = 0$) is able to perform locomotion via nonreciprocal motions of the beads\az{,} resulting in a directed displacement of the center of mass (COM) position $x_\mathrm{c}=(x_1+x_2+x_3)/3$ \cite{Najafi2004}. We choose as basic units the bead radius $R$, the viscosity $\eta$ and the maximum force on a bead $F_0$ such that $|F_i|<F_0$.
Hence the unit of time is $T_0=\eta R^2/F_0$.
In previous studies of this model either the forces or the linearly connected bead velocities $v_i(t)$ have been prescribed via a periodic, nonreciprocal motion pattern \cite{Najafi2004,Earl2007,Golestanian2008}. Alternatively a Q-learning procedure \cite{Tsang2018} has been applied \az{(see also Discussion section)}. 
In our ML approach the swimmer does not follow a
prescribed motion but is able to move forward after sufficiently
long training and to respond to chemical fields autonomously
by a continuous change of the arm lengths. 

\subsection*{Phase one: Learning unidirectional locomotion}

We start by demonstrating that a microswimmer 
is able to learn swimming
in the absence of a chemical field
with the help of a simple genetic algorithm. This is achieved by applying RL \cite{Sutton1998} \az{using} a reward scheme   \az{which optimizes} the microswimmer's strategy of locomotion along a prescribed direction within a viscous fluid environment.

\begin{figure}
\includegraphics[width=\columnwidth]{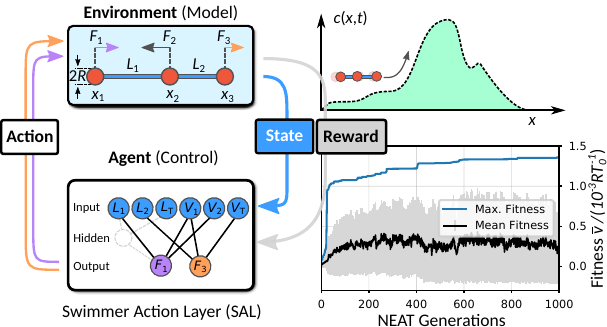}
\caption{Schematic representation of the RL cycle for a   three-bead swimmer moving in a viscous environment (top left) controlled by an ANN-based agent (bottom left). Reward is maximized during training and is granted either for unidirectional locomotion -- phase one -- or for chemotaxis (top right) -- phase two. 
Bottom right: typical NEAT training curves showing the maximum (blue), the mean (black), and the standard deviation (gray) of the fitness (i.e.\ of the cumulative reward) of successive  NEAT generations
each 
covering \bh{200} neural networks
when learning unidirectional locomotion.
}
\label{Fig:1}
\end{figure}

RL algorithms are designed to optimize the \textit{policy} of a so-called  \textit{agent} during training: 
In general, the policy is a highly complex and task-specific quantity that maps the state of an \textit{environment}, i.e.\ everything the agent can perceive \az{(input)}, onto actions which the agent can actively propose \az{(output)} in order to maximize an objective (or reward) function (see Fig.~\ref{Fig:1}). Such rewards might be related to maximize the score of a computer game \cite{Mnih2013}, to minimize the (free) energy when folding proteins \cite{Senior2020}, or -- as in our case -- to maximize the distance that a microswimmer actively moves along a certain direction.

In our approach the agent represents the internal decision making machinery responsible for the deformations of the microswimmer. 
The agent  takes as input (i.e.\ as information it needs to decide about future actions) the state of the environment given by the instantaneous arm lengths $L_1(t)$ and  $L_2(t)$,
\az{and arm velocities $V_i(t)=dL_i(t)/dt$, $i=1,2$.}
In addition 
\az{we use}
the total length $L_{\rm T}(t)=L_1(t)+L_2(t)$, and 
\az{the velocity $V_{\rm T}(t)=V_1(t)+V_2(t)$
as input.}
The arm lengths are normalized by the default length $L_0=10R$ and  subjected to restoring forces acting when $L_1$,$L_2$ are $>1.3L_0$ or $<0.7L_0$ in order to limit the extent of $L_1$ and $L_2$ (see SI Appendix).  
With this information the agent proposes actions which in our case are the forces $F_1(t)$ and  $ F_3(t)$
that determine the dynamics of the swimmer. 
The full hydrodynamic environment, including the three-bead model of the microswimmer, represents the (interactive) environment, whose state is updated 
after the agent has actively proposed its actions (see left part of Fig.~\ref{Fig:1}). In an effort to train unidirectional motion we choose the COM position $x_\mathrm{c}$ of a microswimmer to be maximized 
after a fixed integration time
$T_\mathrm{I}$;
\az{$x_{\rm c}$ thus} represents the cumulative reward of this training process. In this manner we achieve positive reinforcement when the swimmer moves to the right (positive $x$ direction) and negative reinforcement when  it swims to the left (negative $x$ direction).

In \az{order} to approximate the analytically unknown \az{optimum} policy of the microswimmer 
we use ANNs \az{where the output neurons are connected to the input vector,
either directly or through emergent hidden neurons,
using nonlinear activation functions whose arguments depend on the weights of the connections   (see bottom left panel of Fig.~\ref{Fig:1} and Methods).}
In our case the internal structure of the ANN 
(weights and topology)  is successively optimized using the NEAT genetic algorithm 
to maximize the reward
(for details see Methods and SI Appendix).

\begin{figure*} 
\centering
\includegraphics[width=11.4cm, clip=True]{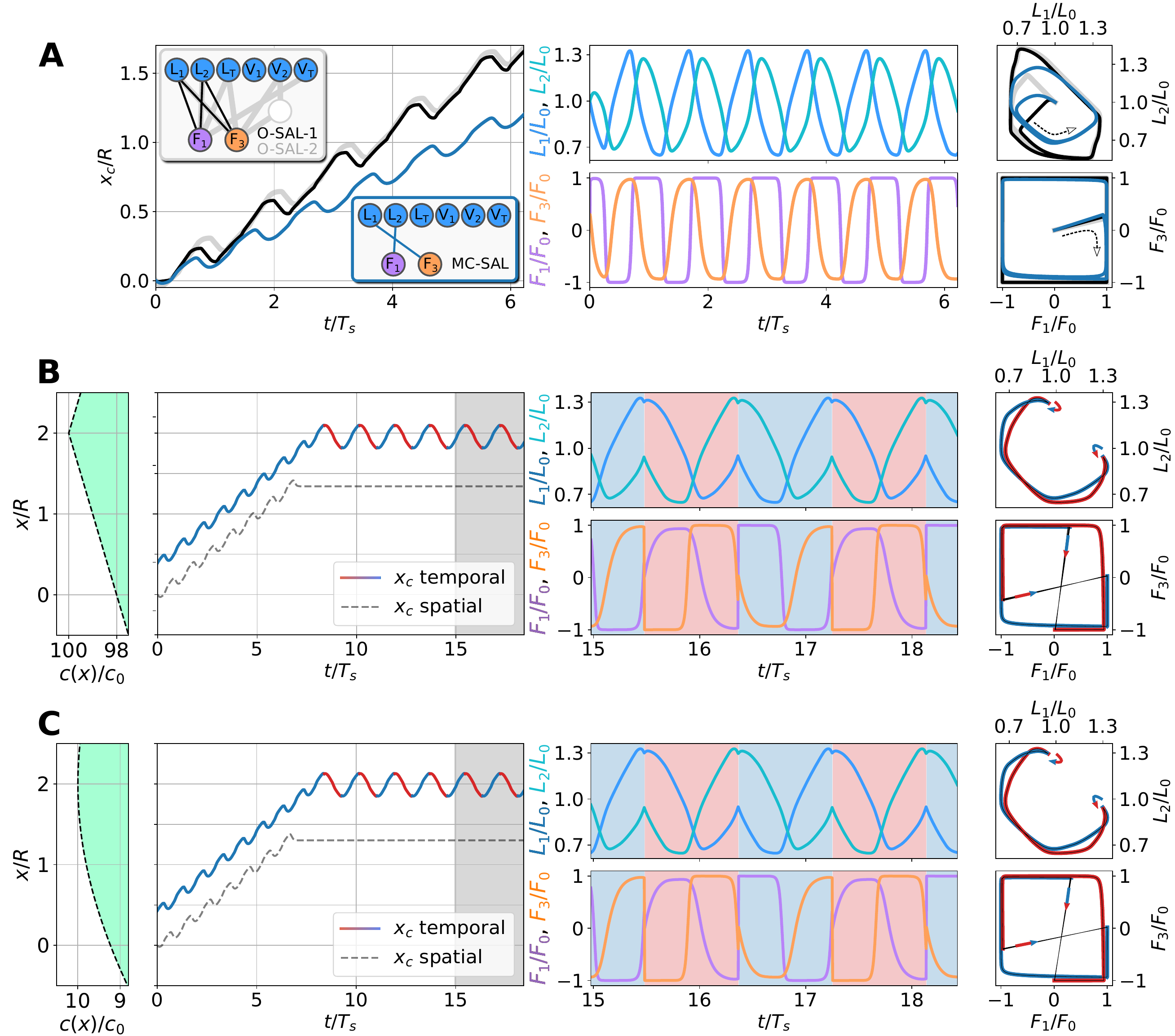} %
\caption{Trajectories of the three-bead swimmer after training. (A) \az{Swimming in the absence of a chemical field. Left:} Time evolution of the  center of mass  $x_\mathrm{c}$ \az{for optimum (O-SAL-1 (black) and O-SAL-2 (gray)) and minimal complexity (MC-SAL) (blue) ANN solutions. Insets show the corresponding topologies.
Time $t$ is shown in units of the MC-SAL stroke period $T_S=217T_0$.
Center: Corresponding arm length solutions  $L_1(t)$ and $L_2(t)$, and  arm forces, $F_1(t)$ and $F_3(t)$,  shown for MC-SAL.} 
\az{Right:} Phase space curves 
($L_1,L_2$) and ($F_1,F_3$) 
\az{for O-SAL-1 (black), O-SAL-2 (gray) und MC-SAL (blue).}
(B) \az{Similar} as in (A), but for an \az{MC-SAL} swimmer in a linear chemical field (see left-most panel), $c(x)=\max(0, a-k|x-x_0|)$ for an amplitude $a=100c_0$, slope $k=c_0/R$, and peak position \bh{$x_0=2R$}, with temporal (red and blue trajectories of \az{$x_{\rm c}$}) and spatial (black dashed trajectory  of \az{$x_{\rm c}$}) chemical gradient sensing (see Fig.~\ref{Fig:3} \az{for ANN solutions}).
Temporal sensing trajectories and phase-space plots are color coded by the currently estimated gradient direction (blue: rightwards, red: leftwards).
\az{The} lengths $L_i(t)$ and forces $F_i(t)$ are shown 
\az{for the time domain highlighted by a gray area in the left trajectory plot.}
\az{Blue} and red background colors correspond to gradient direction estimation (rightwards and leftwards, respectively). Arrows in phase space  indicate locomotive strategy change (gait adaptation) due to gradient estimation (change from rightwards to leftwards locomotion: blue to red, and \textit{vice versa}).
(C) Same as in  (B), but for a swimmer in a Gaussian chemical field (see left-most panel) $c(x)=a\exp(-|x-x_0|^2/2\sigma^2)$ for $a=10c_0$, $\sigma=5R$ and \bh{$x_0=2R$}.}
\label{Fig:2}
\end{figure*}

The training of the swimmer agent is performed over multiple RL \az{steps which correspond to successive NEAT generations. At each step} 
an ensemble of \bh{$N=200$} ANNs (representing one generation) 
controls \az{the swimming gaits of} an ensemble of $N$ independent
microswimmers. 
The cumulative reward $x_\mathrm{c}(T_\mathrm{I})$ is 
evaluated separately for each microswimmer trajectory
defining the \az{\textit{fitness}} \az{$\bar{v}=x_\mathrm{c}(T_\mathrm{I})/T_\mathrm{I}$} of the related ANN-based agent, \az{which is simply the mean swimming velocity (i.e.\ reward per unit time).} 
\az{To start the training, we initialize $N$} ANNs 
where input neurons are 
only sparsely connected to output neurons \az{by using random weights.} 
\az{
The NEAT algorithm then dynamically produces ANN solutions which
differ in number of connections and values of the weights
and may contain
hidden neurons. 
We use the hyperbolic tangent -- tanh(x) -- as output activation functions.}
ANN solutions 
with large fitness values are retained and are preferentially selected for reproduction to form the next generation of ANNs. Thus, good traits of the controlling networks will prevail over time directing thereby the entire ensemble of ANNs to the desired solution. 
\az{In order to capture the possible diversity of genetic pathways we have performed ten independent training runs.}
A  typical \az{evolution of the fitness values of the ANN ensemble}  
is shown in the bottom right panel of Fig.~\ref{Fig:1},
\az{
highlighting the maximum fitness per generation (blue curve)
}
\az{which converges 
to $\bar{v}\rightarrow 1.36\cdot 10^{-3}R/T_0$.}
\az{Similar maximum fitness curves are obtained from the other training runs (SI Appendix Fig.~S1).}
\az{Interestingly, our NEAT training procedure reveals a broad spectrum of network topology solutions (see typical time evolution in Movie S1 and SI Appendix Fig.~S2),
differing in number of connections and hidden neurons. 
Various solutions have high fitness $\bar{v}_{\rm{O}}\approx 1.33-1.36\cdot 10^{-3}R/T_0$,
which we refer to as 
\textit{optimal swimmer action layer} (O-SAL) solutions,
two of them illustrated in the top inset  of Fig.~\ref{Fig:2}A:
}
\az{
The simplest O-SAL solution
does not use any hidden neurons and consist of a sparse architecture 
containing  only four connections (O-SAL-1, thin black connections).
Increasing the number of connections or including hidden neurons only slightly helps to improve the fitness (by $\sim 2 \%$, see also SI Appendix Fig.~S3).
The fittest solution we have found (O-SAL-2, thick gray connections)
uses one hidden neuron and eight connections.
We note that 
more  O-SAL solutions exist, again of different topology but of very similar fitness, 
demonstrating the various possible ANN topologies to obtain maximum fitness 
(SI Appendix Fig.~S4).
The resulting back-and-forth motion of the corresponding swimmer's COM positions $x_{\rm c}(t)$ obtained after training are shown in Fig.~\ref{Fig:2}A(left).
The learned optimum policy describes a square-like shape in $(F_1,F_3)$ action space, 
while the shape of the $(L_1,L_2)$ curve is nontrivial (Fig.~\ref{Fig:2}A(right)).
Again, all O-SAL solutions feature similar trajectories $x_{\rm c}(t)$ and a robust swimming gait (Fig.~\ref{Fig:2}A and SI Appendix Fig.~S4).
}

\az{Strikingly, the algorithm identified intermediate, non-optimum but extremely simple solutions
which can be easily interpreted and consist of as few as two connections  (see also SI Appendix Figs.~S3 and S4).
The best of those 
solutions identified during the NEAT training 
(see bottom  inset of Fig.~\ref{Fig:2}A), 
has still good fitness, $\bar{v}_{MC}=0.95\cdot 10^{-3}R/T_0$,
and we refer to this solution as the \textit{minimal complexity swimmer action layer} (MC-SAL):
$F_1=F_0 \tanh(w_1L_2+b_1)$ and  $F_3=F_0 \tanh(w_2L_1+b_2) $,
with weights $w_1=20.2/L_0$, $w_2=5.7/L_0$, $b_1=-18.6$ and $b_2=-5.4$. 
Here the simple topology, together with the sign and strengths of the weights allow to interpret the occurrence of the phase-shifted periodic output of the arm lengths and the forces (see Fig.~\ref{Fig:2}A, Movie S2 and discussion in SI Appendix).
Finally, alternative yet less efficient minimal complexity strategies are also possible (SI Appendix  Fig.~S4).
}


\subsection*{Phase two: Learning chemotaxis in a constant gradient -- spatial vs. temporal gradient detection}

Now we proceed to the challenging problem of finding a policy which allows the microswimmer to navigate on its own within a complex environment such as a chemical field, $c(x)$ (cf. upper right panel in Fig.~\ref{Fig:1}), and perform positive chemotaxis (i.e., motion towards local maxima of $c(x)$).

We first extend
the agent's perception of
the environment 
such that it is able to sense the field $c(x)$ (which we 
normalize by an arbitrary concentration strength $c_0$) and which we
use as an additional  
input for a more advanced chemotaxis agent. 
We expect that such an agent is able to evaluate the chemical \textit{gradient} $\nabla c(x)$ in order to conditionally control the lengths of its arms 
in a way to steer its motion towards maxima of $c(x)$.
Compared to phase one we propose a slightly more complex cumulative reward scheme for the training phase: we use $r_{\rm c}=\sum_{t_i=1}^{T_\mathrm{I}}[x_\mathrm{c}(t_{i}) - x_\mathrm{c}(t_{i-1})] D(t_i)$ where  $D(t_i)=\textrm{sign}[\nabla c(x_\mathrm{c}(t_{i}))] = \pm 1$ represents the sign of the gradient at instant $t_i$; thus, $r_{\rm c}$ 
measures the total distance that the swimmer moves along an ascending gradient during the total integration time $T_\mathrm{I}$.

\begin{figure}
\includegraphics[width=\columnwidth]{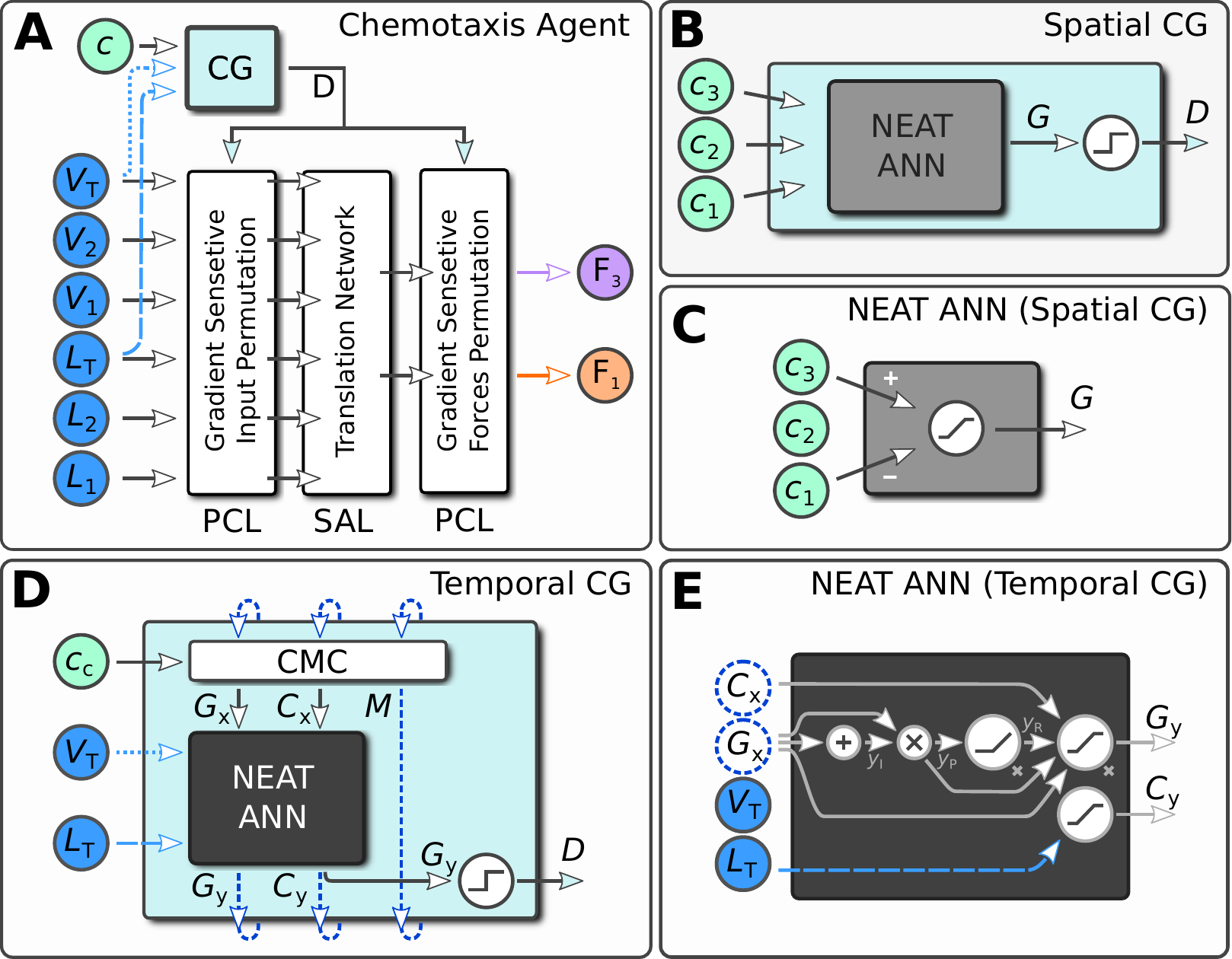}
\caption{
(A) Schematic view of full ANN-based chemotaxis agent.
\az{A chemical field $c(x)$, swimmer arm lengths ($L_1$, $L_2$, $L_\mathrm{T} = L_1 + L_2$), and respective arm velocities ($V_1$, $V_2$, $V_\mathrm{T}$) are used as input.}
By measuring the chemical gradient  through the CG-block the swimmer controls the forces $F_1$ and $F_3$ in order to perform directed locomotion towards an ascending gradient of $c(x)$.
Directed locomotion is split
into two permutation control layers (PCL) 
which permute  input and  output of the swimmer action layer (SAL) (see insets of Fig.~\ref{Fig:2}A) according to a predicted sign $D$ of the chemical gradient. 
The prediction of $D$ by the CG-block (cyan) can be performed either by directly measuring $D=\mathrm{sign}(\nabla c(x_\mathrm{c}))$, or by (B) spatial resolution of the chemical field, or by (D) temporal sensing at the center of mass position $x_\mathrm{c}$. 
\az{The respective solutions for the ANNs (dark gray and gray) found by NEAT are shown in (C) and (E).}
}
\label{Fig:3}
\end{figure}

Prior to applying any RL scheme we decompose the problem of chemotaxis  into two 
tasks: first, we require a mechanism which allows the agent to discern the direction $D$ of the gradient (i.e. $D=1$ for ascending or $D=-1$ for descending); we introduce this tool as a \textit{chemical gradient \az{(CG)} block} in the ANN of the chemotaxis agent (see Fig.~\ref{Fig:3}A) as described below.
Second, we identify a pure locomotive part of the agent which can be rooted on already acquired skills -- i.e.\ the unidirectional motion learned in phase one (and covered by the above mentioned SAL \az{solutions}) -- and on the inherent symmetries of the swimmer model: swimming to the left and swimming to the right are symmetric operations. Based on the actual value of $D$, conditional directional motion (i.e., either to the left or to the right) can be induced by introducing two permutation control layers (PCLs) to the ANN (see Fig.~\ref{Fig:3}A, \az{and} SI Appendix
\az{for details}).


In order to obtain chemotaxis strategies using NEAT, the  remaining task is to identify a (potentially recurrent) ANN structure for the chemical gradient block (Fig.~\ref{Fig:3}A), i.e. an ANN which is able to predict the sign $D$ of the chemical gradient. For this purpose we have considered three different methods which allow the microswimmer to sense $\nabla c(x)$: first, we assume that the chemotaxis agent can directly measure the sign of the  gradient at its COM position $x_\mathrm{c}(t)$: here $D$ is automatically known. Second, we allow the swimmer to simultaneously evaluate the chemical fields \az{$c_i(t)$} at the bead positions $x_i(t)$ to predict the \az{sign of the gradient via
\bh{$D=\mathrm{sign}(G)$} 
from the output $G$ of the ANN (Fig.~\ref{Fig:3}B)\bh{,}  
determined by NEAT during training (see below)}.  
Third, in an effort to model  \textit{temporal} sensing of chemical gradients, which is relevant for bacterial chemotaxis, we consider recurrent ANNs \az{(Fig.~\ref{Fig:3}D)}.
\az{In this case, we explicitly provide the CG agent with}
\az{inputs that describe the}
\az{internal, physiological state (total arm length $L_T$ and velocity $V_T$), as well as with the chemical field at the COM position $c_c=c(x_c)$
at each instance of time $t_i$.
To train the CG agent 
we subdivide its task 
into a block which estimates the gradient, and into another block that controls an internal memory of the chemical field (i.e., the chemical memory control (CMC) cell).
The latter is inspired by the well-known long short-term memory (LSTM) cell \cite{Hochreiter1997,Staudmeyer2019}.}


\az{The first block is trained using the NEAT algorithm: it takes as input $L_T(t_i)$ and $V_T(t_i)$ as well as two recurrent variables $C_x(t_i)$ and $G_x(t_i)$ and maps this information onto a control output $C_y(t_i)$ and an estimated value of the instantaneous chemical gradient $G_y(t_i)$,
both to be processed by the CMC cell in the next time step.
The CMC cell temporally feeds back $G_y$ as input to the NEAT ANN as $G_x(t_{i})=G_y(t_{i-1})$.
Furthermore, 
the CMC cell controls via 
the binary variable $\beta=\Theta(C_y(t_{i-1}))= \{ 0,1 \}$ \bh{(with $\Theta(\cdot)$ the Heaviside function)}
the state of an internal memory $M(t_{i}) = (1-\beta)\,M(t_{i-1})+\beta\,c_c(t_{i})/c_0$
and the state of the NEAT ANN input $C_x(t_{i}) = (1-\beta)\, C_y(t_{i-1})+\beta\,(c_c(t_{i})/c_0-M(t_{i-1}))$.
In that way the CG agent can actively control the time interval between successive measurements: 
an update of $M(t_{i})$ is performed whenever $C_y(t_{i-1})>0$, otherwise $M(t_{i-1})$ is maintained over time.
Notably, the chemical field input of the CG agent is directly forwarded to the CMC cell and the trained NEAT ANN operates on \bh{time-delayed} gradients rather than directly on the values of the chemical field $c_c$: whenever $C_y(t_{i-1})>0$ the CMC cell explicitly provides temporal gradient information $(c_c(t_{i})/c_0-M(t_{i-1}))$ to the NEAT ANN via $C_x(t_{i})$, otherwise feeds back $C_y(t_{i-1})$.
Eventually, the output of the temporal CG agent is \bh{$D=\mathrm{sign}(G_y)$}.} 

For temporal and spatial gradient sensing (Fig.~\ref{Fig:3}B,D)
training is necessary.
For simplicity, we train the swimmer on a piece-wise linear field, $c(x)=\max(0, a-k|x-x_0|)$, with amplitude $a$ and slope $k$ \az{using the MC-SAL solution obtained in phase one} (see Movies S3 and S4, SI Appendix and Figs.~S9 and S12 
for details).

Both for spatial and temporal sensing methods
the resulting ANNs
are strikingly simple  and their topology can be well interpreted: the NEAT ANN solution for spatial sensing\az{, shown in Fig.~\ref{Fig:3}C,} only requires a single neuron which predicts 
\bh{$D(t)\approx \mathrm{sign}(c_3(t) - c_1(t))$} 
(see 
SI Appendix for details).
During training of the temporal gradient-sensing ANN we determine the precise way how \bh{the} output \bh{signals} of the ANN \bh{ $C_y(t_i)$ and $G_y(t_i)$ are} used as recurrent input \bh{signals} in the next time step and how \bh{$C_y(t_i)$} controls the way the chemical memory is updated. 
The solution for temporal sensing \az{is shown Fig.~\ref{Fig:3}E.} 
\bh{
    The NEAT evolved ANN has learned to exploit the periodically changing total arm-length $L_T(t_i)$ as a pacemaker for inducing chemical memory updates via $C_y(t_i)>0$
    \az{whenever} $L_T(t_i)\gtrsim 2.23L_0$ (see SI Appendix Figs.~S15 to S17).
    The functional form of how to predict the time-delayed chemical gradient via $G_y(t_i)$ is more involved:
    First, the recurrent input of $G_x(t_i)$ is bypassed via $y_\mathrm{I}\propto G_x(t_i)$ in a single identity neuron (labeled by the $\oplus$ symbol), which is then multiplied again with $G_x(t_i)$ in an identity product neuron 
    (labeled by the $\otimes$ symbol) with output $y_\mathrm{P}\propto G_x^2(t_i)$.
    This squared recurrent signal $y_\mathrm{P}$ is then transformed by a \az{rectified linear unit (``relu'', see SI Appendix)} with output $y_R$.
    Eventually, the output neuron $G_y(t_i)$ multiplies the weighted signals $G_x(t_i)$, $y_\mathrm{R}$, $y_\mathrm{P}$ and $C_x(t_i)$; 
    the output then represents the estimate for the chemical gradient.
    Noteworthy, $C_x(t_i)$ plays a non-trivial, two-fold role in the gradient estimate: whenever $C_y(t_{i-1})<0$, $C_x(t_i)$ is a function of the former total arm-length $L_\mathrm{T}(t_{i-1})$, otherwise it represents the delayed chemical gradient $C_x(t_i)=c_c(t_i)/c_0-M(t_{i-1})$ between two measurement steps.}
\bh{In that way, the} temporal gradient-sensing ANN enables the \az{CG} agent to correlate its direction of propagation with the gradient of a chemical field.
\bh{
Numerical details on the weights and biases, and further interpretation of the ANN solution depicted in Fig.~\ref{Fig:3}E are} provided in the SI Appendix.




In Figs.~\ref{Fig:2}B and C we present 
typical trajectories after successful training
obtained for chemical fields of piece-wise linear shape  and of Gaussian shape, 
respectively. In both cases the swimmer -- controlled by spatial sensing -- suddenly stops as soon as its COM position $x_\mathrm{c}$ is reasonably close to the maximum $x_0$ of the chemical field (see also \az{Movie~S7}).  In contrast, the swimmer controlled by temporal sensing performs oscillations around $x_0$ due to its \bh{time-delayed} measurements of the chemical field and its internal, recurrent processes (see Fig.~\ref{Fig:3}B and \az{Movies S5 and S6}). 

We observe that the ANNs of both spatial and temporal sensing methods are able to generalize their capability to predict the chemical gradient over a much wider range of parameters (i.e., amplitude $a$ and slope $k$ of a chemical field) than they were originally trained on (see \az{Fig.~\ref{Fig:2}}, SI Appendix and \az{Figs.~S9 and S14.})  

\begin{figure}
  \includegraphics[width=\columnwidth]{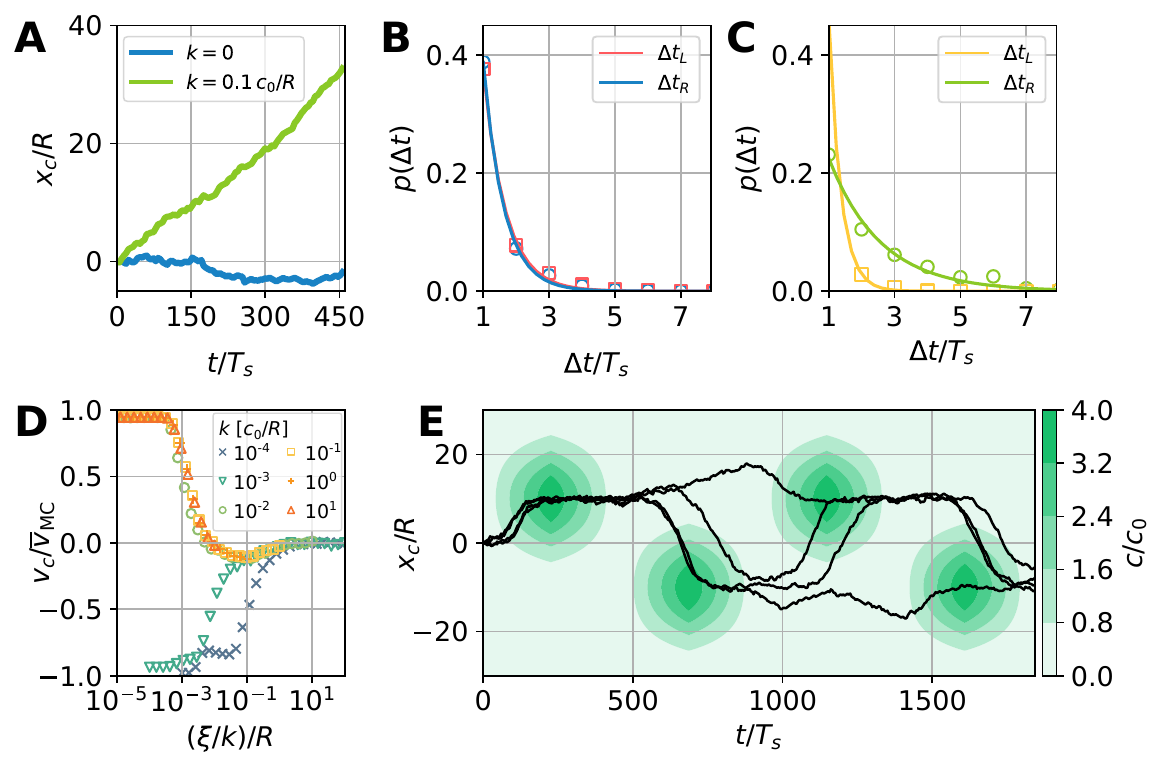}
\caption{
Stochastic microswimmer dynamics from noisy memory readings for noise level \bh{$\xi=2\cdot10^{-4}c_0$}. 
(A) Sample trajectories in the absence (\bh{blue}) and in the presence (green) of a linear chemical field.
(B,C) Run time distributions for moving the field upwards ($\Delta t_R$) and downwards ($\Delta t_L$) in the absence (B) and in the presence (C) of a field.
\az{Note the run time axis starts at the minimal possible run time $\Delta t/T_0=1$ because of the used discretization.}
\mh{(D) Chemotactic drift velocity $v_c$ as a function of noise-to-signal ratio \bh{$\xi/(kR)$} for different values of gradient steepness $k$. 
Each data point corresponds to a simulation time of $10^6 T_0$} 
(E) Sample trajectories in  time-dependent Gaussian profiles $c(x,t)$ (see color bar)
centered at \mh{$x_{0\pm}=\pm 10R$}
of width \mh{$\sigma=8R$} and height \mh{$a=4c_0$}, and modulated
with  period \mh{$T=461 T_\mathrm{S}$.}} 
\label{Fig:4}
\end{figure}

\subsection*{Emergent run-reverse motion from noisy memory readings}
Realistic  chemotactic pathways are always influenced by thermal noise.
In our implementation 
we apply stochastic memory readings of the CMC cell for the temporally sensing swimmer, mimicking the fact that the chemotactic signal cannot be detected perfectly. 
In this spirit the swimmer measures a field,
\bh{$M(t)=(c(x_\mathrm{c}(t)) + \delta c)/c_0$, $\delta c$} being a normal distributed random number with zero mean and standard deviation $\xi$ which sets the strength of the noise. We apply this feature to an ensemble of 100 non-interacting microswimmers moving in a constant chemical gradient $c(x)=kx$
\az{but which have learned chemotaxis in the absence of noise in phase two.}
Strikingly, a \az{1D} run-and-tumble \az{(run-and-reverse)} motion emerges naturally, even in the absence of a chemical field ($k=0$).
In Fig.~\ref{Fig:4}A we present typical trajectories both in the absence and in the presence ($k=0.1c_0/R$) of a chemical field (see also \bh{Movie~S8}). These  trajectories consist of segments of rightward motion (over run times $\Delta t_{\rm R}$), alternating with segments of leftward motion ($\Delta t_{\rm L}$). The stochastic nature of the underlying process 
leads to approximately exponentially distributed run times, $\sim e^{-\Delta t_R/\tau_R}$ and $\sim e^{-\Delta t_L/\tau_L}$, following thus a similar behaviour as the one measured for  microorganisms \cite{Berg1972,Polin2009,Theves2013}.
As expected, in the absence of a field $\tau_R\approx \tau_L$ (Fig.~\ref{Fig:4}B). In the presence of a field the swimmer exhibits a tendency for longer run times moving the gradient upwards ($\tau_R>\tau_L$) (Fig.~\ref{Fig:4}C).

\az{In general, the chemotactic performance,
 quantified by the mean net chemotactic drift velocity $v_{\rm c}$ (i.e.\ mean swimmer velocity), 
depends on the gradient steepness $k$ and is strongly influenced by the noise level $\xi$
as shown in Fig.~\ref{Fig:4}D:
As expected, for very small noise the motion is almost ballistic, $v_{\rm c} \rightarrow \bar{v}_{MC}$,
while biased run-and-reverse motion ($0<v_{\rm c}  <\bar{v}_{MC}$) emerges for larger noise.
Interestingly, for different values of $k$, this can be quantified by the noise-to-signal ratio \bh{$\xi/(kR)$} leading to
a universal chemotactic behavior for a large range of $k$ values (see also SI Appendix Fig.~S21).
Note that there exists a noise-to-signal regime where the chemotactic velocity becomes negative due to the small bias
of the microswimmer obtained during training (see SI Appendix).}


\mh{Run-and-reverse behavior depends on the values of the chemical field $c(t_j)$ and \bh{of the} internal memory $M(t_i)$ at two  \bh{distinct}
points in time $t_i < t_j$ the microswimmer chooses to perform \bh{successive} measurement\az{s}.
If the noise 
$\delta c$ 
dominates in the swimmer’s input $C_x(t_j) = (c(t_j) - c(t_i) + \delta c)/c_0$, the swimmer is unable to correctly determine the \bh{time-delayed} chemical gradient and moves erratically. 
This happens \bh{either} if measurements are performed too frequently or if the noise-to-signal ratio is above a critical value of  \bh{$\xi/(kR)\gtrsim10^{-2}$} (see \bh{Fig.~\ref{Fig:4}D}).
For a detailed account of how and when our solution performs a measurement of the chemical field see SI Appendix Fig. S18 and S19.}



\subsection*{Chemotaxis in time-dependent chemical fields}
Eventually we 
study the dynamics of 
temporal gradient sensing microswimmers which perform noisy memory readings  in a more complicated, time-dependent chemical environment.
Notably, the microswimmers
have solely been trained in a constant chemical gradient as described in phase two.
We now use time-dependent chemical fields of the form
${c(x,t) = h_+(t)c_+(x) + h_-(t)c_-(x)}$ where $c_{\pm}(x)$ are of Gaussian shape with maximum height $a$ and centered at
\az{peak positions} $x_{0\pm}$.
The peak amplitudes are modulated via $h_\pm(t)=\sum_{i=0} \mathrm{max}[(1 - |4(i-t/T)\pm1|), 0]$ with period $T$,
see contour plot in \bh{Fig.~\ref{Fig:4}E}
where we also
show typical microswimmer trajectories.
Swimmers may explore consecutive peaks by \textit{hopping} between chemical sources of $c_+$ and $c_-$, or may miss peaks by residing in the vicinity of the previously visited chemical source.
Thus, the actual swimming paths strongly depend on prior decisions of the chemotaxis agent.
In field-free regions microswimmers perform \az{approximately} unbiased 
run-and-reverse strategies and they employ positive chemotaxis 
in regions featuring chemical gradients.
Hence the combination of chemotactic response and noise enables useful foraging strategies in time-dependent fields.

\section*{Discussion}
We modeled the
response of a simple microswimmer to a viscous and chemical environment using the NEAT genetic algorithm
to construct ANNs
which describe the internal decision making machinery coupled to the motion of two arms.
First our model microswimmer learned
to swim in the absence of a chemical field in a \textit{``1 Step Back, 2 Steps Forward''} motion as it appears, for example, for the swimming pattern of the algae \textit{Chlamydomonas}.


In contrast to a recently used Q-learning approach
which uses a very limited action space \cite{Tsang2018},
we  allow continuous changes of the microswimmer's shape and thus permit high flexibility in exploring many different swimming gaits during training. 
\az{This feature allowed us to find optimum swimming policies where the forces on the beads are limited, in contrast to fixing arm velocities (see also SI Appendix).}
Furthermore, the NEAT algorithm has
created surprisingly
simple ANNs which we were able to fully understand and interpret, in contrast to often used complex deep neural networks \cite{Lillicrap2015,Gu2019,Schulman2017,Watkins1992} or the lookup table like Q-learning algorithm \cite{Watkins1992}.

We used biologically relevant chemotactic sensing strategies, namely spatial gradient sensing usually performed by slow-moving eukaryotic cells, and temporal gradient sensing performed by fast swimming bacteria.
We used the latter to explore the influence of a single noisy channel, namely for the reading of the value of the chemical concentration, on the chemotactic response.
Interestingly, we identified  \az{for different values of gradient steepness a broad range of noise levels} for a run-and-\az{reverse} type of dynamics with 
exponentially distributed run times
\az{which can be scaled onto a master curve using the noise-to-signal ratio of the chemical field measurement.}
\az{However, 
this behavior depends on the specific network solution obtained during training in phase two (see also SI Appendix Fig.~S21).}
Indeed for real existing signal sensing mechanisms in microorganisms the role of the noise and the precision of signal detection is an active field of research,
see e.g.~\az{\cite{Berg1977,TenWolde2016a}}.

The run-and-\az{reverse} behavior in our system is an emergent feature which sustains in the absence of a chemical field (as observed, for example, for swimming bacteria)
without explicitly challenging the microswimmer to exploit search strategies in the absence of a  field during training.
From an evolutionary point of view it makes sense that bacteria have \textit{learned} this behavior in complex chemical environments. 
We also find that individual microswimmers performing run-and-\az{reverse} motion may show a small bias to the left or to the right even in the absence of a field due to the stochastic nature of the genetic optimization \az{(see also SI Appendix Figs.~S14 and S21)}. 

The question how single cells make decisions which affect their motion in their environment is an active field of research \cite{Balazsi2011,Bowsher2014,Tang2018,Tripathi2020}.
For example, 
bacteria, protists, plants, and fungi make decisions without using neurons but rather employ a complex chemotactic signaling network \cite{Reid2015}.
On the other hand, small multicellular organisms such as the worm \textit{C.~elegans} use only a small number of neurons in order to move and perform chemotaxis \cite{Jarrell2012,Itskovits2018}.
Our approach therefore offers new tools in order to investigate possible  architectures,  functionalities and the necessary level of complexity of sensing and motor neurons coupled to muscle movement  \textit{in-silico} by  evolutionary 
developed ANNs.
In the future our work can be extended to more specific microswimmers moving in two or three dimensions, in order to
extract the necessary complexity of the decision making machinery 
used for chemotaxis, mechanosensing, or even more complex
 behavioral responses such as reproduction.

 \section*{Methods}
\subsection*{Artificial Neural Networks (ANNs)}
An ANN is a set of interconnected artificial neurons which collect weighted signals (either from external sources or from other neurons) and create and redistribute output signals generated by a nonlinear activation function \cite{Goodfellow2016} (see SI Appendix for details). In that way an ANN can process information in an efficient and 
flexible way: by adjusting the weights and biases of connections between different neurons or by adjusting the network topology ANNs can be trained to map network input to output signals thereby realizing task specific operations which are often too complicated to be implemented manually \cite{Baker1998}.
\subsection*{NEAT algorithm}
\textit{NeuroEvolution of Augmented Topologies} (NEAT) \cite{Stanley2002} is a genetic algorithm designed for constructing neural networks. In contrast to most learning algorithms it does not only optimize the weights of an ANN (in an effort to optimize a so-called target function), but, moreover, generates the weights and the topology of the ANN simultaneously (see SI Appendix for details). This process is guided by the principle of \textit{complexification} \cite{Stanley2002}: starting from a minimal design of the ANN, the algorithm will gradually add or remove nodes and connecting neurons with certain probabilities according the evolutionary process (schematically depicted by the \bh{gray} dashed lines in the bottom left panel of Fig.~\ref{Fig:1}), in order to keep the resulting network as simple and sparse as possible. The resulting ANNs of minimal complexity can then be used to employ the target task, even for situations that the ANNs never have explicitly experienced during training.

\section*{acknowledgements}
B.H. acknowledges a DOC Fellowship of the Austrian Academy of Sciences.
B.H. and G.K. acknowledge financial support by E-CAM, an e-infrastructure center of excellence for software, training, and consultancy in simulation and modeling funded by the EU (Project no. 676531).
A.Z.\ acknowledges funding from the Austrian
Science Fund (FWF) through a Lise-Meitner Fellowship
(Grant No. M 2458-N36).
The computational results presented have been achieved using the Vienna Scientific Cluster.
Helpful discussions with Dr.~Andreas Singraber (Vienna) on neural networks are gratefully acknowledged.


\begin{thebibliography}{57}%
\makeatletter
\providecommand \@ifxundefined [1]{%
 \@ifx{#1\undefined}
}%
\providecommand \@ifnum [1]{%
 \ifnum #1\expandafter \@firstoftwo
 \else \expandafter \@secondoftwo
 \fi
}%
\providecommand \@ifx [1]{%
 \ifx #1\expandafter \@firstoftwo
 \else \expandafter \@secondoftwo
 \fi
}%
\providecommand \natexlab [1]{#1}%
\providecommand \enquote  [1]{``#1''}%
\providecommand \bibnamefont  [1]{#1}%
\providecommand \bibfnamefont [1]{#1}%
\providecommand \citenamefont [1]{#1}%
\providecommand \href@noop [0]{\@secondoftwo}%
\providecommand \href [0]{\begingroup \@sanitize@url \@href}%
\providecommand \@href[1]{\@@startlink{#1}\@@href}%
\providecommand \@@href[1]{\endgroup#1\@@endlink}%
\providecommand \@sanitize@url [0]{\catcode `\\12\catcode `\$12\catcode
  `\&12\catcode `\#12\catcode `\^12\catcode `\_12\catcode `\%12\relax}%
\providecommand \@@startlink[1]{}%
\providecommand \@@endlink[0]{}%
\providecommand \url  [0]{\begingroup\@sanitize@url \@url }%
\providecommand \@url [1]{\endgroup\@href {#1}{\urlprefix }}%
\providecommand \urlprefix  [0]{URL }%
\providecommand \Eprint [0]{\href }%
\providecommand \doibase [0]{http://dx.doi.org/}%
\providecommand \selectlanguage [0]{\@gobble}%
\providecommand \bibinfo  [0]{\@secondoftwo}%
\providecommand \bibfield  [0]{\@secondoftwo}%
\providecommand \translation [1]{[#1]}%
\providecommand \BibitemOpen [0]{}%
\providecommand \bibitemStop [0]{}%
\providecommand \bibitemNoStop [0]{.\EOS\space}%
\providecommand \EOS [0]{\spacefactor3000\relax}%
\providecommand \BibitemShut  [1]{\csname bibitem#1\endcsname}%
\let\auto@bib@innerbib\@empty
\bibitem [{\citenamefont {Purcell}(1977)}]{Purcell1977}%
  \BibitemOpen
  \bibfield  {author} {\bibinfo {author} {\bibfnamefont {E.~M.}\ \bibnamefont
  {Purcell}},\ }\href@noop {} {\bibfield  {journal} {\bibinfo  {journal} {Am.
  J. Phys.}\ }\textbf {\bibinfo {volume} {45}},\ \bibinfo {pages} {3} (\bibinfo
  {year} {1977})}\BibitemShut {NoStop}%
\bibitem [{\citenamefont {Lauga}\ and\ \citenamefont
  {Powers}(2009)}]{Lauga2009a}%
  \BibitemOpen
  \bibfield  {author} {\bibinfo {author} {\bibfnamefont {E.}~\bibnamefont
  {Lauga}}\ and\ \bibinfo {author} {\bibfnamefont {T.~R.}\ \bibnamefont
  {Powers}},\ }\href {\doibase 10.1088/0034-4885/72/9/096601} {\bibfield
  {journal} {\bibinfo  {journal} {Rep. Prog. Phys.}\ }\textbf {\bibinfo
  {volume} {72}},\ \bibinfo {pages} {096601} (\bibinfo {year} {2009})},\
  \Eprint {http://arxiv.org/abs/0812.2887} {arXiv:0812.2887} \BibitemShut
  {NoStop}%
\bibitem [{\citenamefont {Elgeti}\ \emph {et~al.}(2015)\citenamefont {Elgeti},
  \citenamefont {Winkler},\ and\ \citenamefont {Gompper}}]{Elgeti2015}%
  \BibitemOpen
  \bibfield  {author} {\bibinfo {author} {\bibfnamefont {J.}~\bibnamefont
  {Elgeti}}, \bibinfo {author} {\bibfnamefont {R.~G.}\ \bibnamefont {Winkler}},
  \ and\ \bibinfo {author} {\bibfnamefont {G.}~\bibnamefont {Gompper}},\ }\href
  {\doibase 10.1088/0034-4885/78/5/056601} {\bibfield  {journal} {\bibinfo
  {journal} {Rep. Prog. Phys.}\ }\textbf {\bibinfo {volume} {78}},\ \bibinfo
  {pages} {056601} (\bibinfo {year} {2015})},\ \Eprint
  {http://arxiv.org/abs/1412.2692} {arXiv:1412.2692} \BibitemShut {NoStop}%
\bibitem [{\citenamefont {Z{\"{o}}ttl}\ and\ \citenamefont
  {Stark}(2016)}]{Zoettl2016}%
  \BibitemOpen
  \bibfield  {author} {\bibinfo {author} {\bibfnamefont {A.}~\bibnamefont
  {Z{\"{o}}ttl}}\ and\ \bibinfo {author} {\bibfnamefont {H.}~\bibnamefont
  {Stark}},\ }\href {\doibase 10.1088/0953-8984/28/25/253001} {\bibfield
  {journal} {\bibinfo  {journal} {J. Phys.: Condens. Matter}\ }\textbf
  {\bibinfo {volume} {28}},\ \bibinfo {pages} {253001} (\bibinfo {year}
  {2016})},\ \Eprint {http://arxiv.org/abs/1601.06643} {arXiv:1601.06643}
  \BibitemShut {NoStop}%
\bibitem [{\citenamefont {Berg}\ and\ \citenamefont {Brown}(1972)}]{Berg1972}%
  \BibitemOpen
  \bibfield  {author} {\bibinfo {author} {\bibfnamefont {H.~C.}\ \bibnamefont
  {Berg}}\ and\ \bibinfo {author} {\bibfnamefont {D.~A.}\ \bibnamefont
  {Brown}},\ }\href {\doibase 10.1038/239500a0} {\bibfield  {journal} {\bibinfo
   {journal} {Nature}\ }\textbf {\bibinfo {volume} {239}},\ \bibinfo {pages}
  {500} (\bibinfo {year} {1972})}\BibitemShut {NoStop}%
\bibitem [{\citenamefont {Lauga}(2016)}]{Lauga2016}%
  \BibitemOpen
  \bibfield  {author} {\bibinfo {author} {\bibfnamefont {E.}~\bibnamefont
  {Lauga}},\ }\href {\doibase 10.1146/annurev-fluid-122414-034606} {\bibfield
  {journal} {\bibinfo  {journal} {Annu. Rev. Fluid Mech.}\ }\textbf {\bibinfo
  {volume} {48}},\ \bibinfo {pages} {105} (\bibinfo {year} {2016})},\ \Eprint
  {http://arxiv.org/abs/1509.02184} {arXiv:1509.02184} \BibitemShut {NoStop}%
\bibitem [{\citenamefont {Polin}\ \emph {et~al.}(2009)\citenamefont {Polin},
  \citenamefont {Tuval}, \citenamefont {Drescher}, \citenamefont {Gollub},\
  and\ \citenamefont {Goldstein}}]{Polin2009}%
  \BibitemOpen
  \bibfield  {author} {\bibinfo {author} {\bibfnamefont {M.}~\bibnamefont
  {Polin}}, \bibinfo {author} {\bibfnamefont {I.}~\bibnamefont {Tuval}},
  \bibinfo {author} {\bibfnamefont {K.}~\bibnamefont {Drescher}}, \bibinfo
  {author} {\bibfnamefont {J.~P.}\ \bibnamefont {Gollub}}, \ and\ \bibinfo
  {author} {\bibfnamefont {R.~E.}\ \bibnamefont {Goldstein}},\ }\href {\doibase
  10.1126/science.1172667} {\bibfield  {journal} {\bibinfo  {journal}
  {Science}\ }\textbf {\bibinfo {volume} {487}},\ \bibinfo {pages} {487}
  (\bibinfo {year} {2009})}\BibitemShut {NoStop}%
\bibitem [{\citenamefont {Clark}\ and\ \citenamefont
  {Grant}(2005)}]{Clark2005}%
  \BibitemOpen
  \bibfield  {author} {\bibinfo {author} {\bibfnamefont {D.~A.}\ \bibnamefont
  {Clark}}\ and\ \bibinfo {author} {\bibfnamefont {L.~C.}\ \bibnamefont
  {Grant}},\ }\href {\doibase 10.1073/pnas.0407659102} {\bibfield  {journal}
  {\bibinfo  {journal} {Proceedings of the National Academy of Sciences of the
  United States of America}\ }\textbf {\bibinfo {volume} {102}},\ \bibinfo
  {pages} {9150} (\bibinfo {year} {2005})}\BibitemShut {NoStop}%
\bibitem [{\citenamefont {Celani}\ and\ \citenamefont
  {Vergassola}(2010)}]{Celani2010}%
  \BibitemOpen
  \bibfield  {author} {\bibinfo {author} {\bibfnamefont {A.}~\bibnamefont
  {Celani}}\ and\ \bibinfo {author} {\bibfnamefont {M.}~\bibnamefont
  {Vergassola}},\ }\href {\doibase 10.1073/pnas.0909673107} {\bibfield
  {journal} {\bibinfo  {journal} {Proceedings of the National Academy of
  Sciences of the United States of America}\ }\textbf {\bibinfo {volume}
  {107}},\ \bibinfo {pages} {1391} (\bibinfo {year} {2010})}\BibitemShut
  {NoStop}%
\bibitem [{\citenamefont {Taktikos}\ \emph {et~al.}(2013)\citenamefont
  {Taktikos}, \citenamefont {Stark},\ and\ \citenamefont
  {Zaburdaev}}]{Taktikos2013}%
  \BibitemOpen
  \bibfield  {author} {\bibinfo {author} {\bibfnamefont {J.}~\bibnamefont
  {Taktikos}}, \bibinfo {author} {\bibfnamefont {H.}~\bibnamefont {Stark}}, \
  and\ \bibinfo {author} {\bibfnamefont {V.}~\bibnamefont {Zaburdaev}},\ }\href
  {\doibase 10.1371/journal.pone.0081936} {\bibfield  {journal} {\bibinfo
  {journal} {PLoS ONE}\ }\textbf {\bibinfo {volume} {8}} (\bibinfo {year}
  {2013}),\ 10.1371/journal.pone.0081936}\BibitemShut {NoStop}%
\bibitem [{\citenamefont {Alirezaeizanjani}\ \emph {et~al.}(2020)\citenamefont
  {Alirezaeizanjani}, \citenamefont {Gro{\ss}mann}, \citenamefont {Pfeifer},
  \citenamefont {Hintsche},\ and\ \citenamefont {Beta}}]{Alirezaeizanjani2020}%
  \BibitemOpen
  \bibfield  {author} {\bibinfo {author} {\bibfnamefont {Z.}~\bibnamefont
  {Alirezaeizanjani}}, \bibinfo {author} {\bibfnamefont {R.}~\bibnamefont
  {Gro{\ss}mann}}, \bibinfo {author} {\bibfnamefont {V.}~\bibnamefont
  {Pfeifer}}, \bibinfo {author} {\bibfnamefont {M.}~\bibnamefont {Hintsche}}, \
  and\ \bibinfo {author} {\bibfnamefont {C.}~\bibnamefont {Beta}},\ }\href
  {\doibase 10.1126/sciadv.aaz6153} {\bibfield  {journal} {\bibinfo  {journal}
  {Science Advances}\ }\textbf {\bibinfo {volume} {6}} (\bibinfo {year}
  {2020}),\ 10.1126/sciadv.aaz6153}\BibitemShut {NoStop}%
\bibitem [{\citenamefont {Swaney}\ \emph {et~al.}(2010)\citenamefont {Swaney},
  \citenamefont {Huang},\ and\ \citenamefont {Devreotes}}]{Swaney2010}%
  \BibitemOpen
  \bibfield  {author} {\bibinfo {author} {\bibfnamefont {K.~F.}\ \bibnamefont
  {Swaney}}, \bibinfo {author} {\bibfnamefont {C.-H.}\ \bibnamefont {Huang}}, \
  and\ \bibinfo {author} {\bibfnamefont {P.~N.}\ \bibnamefont {Devreotes}},\
  }\href {\doibase 10.1146/annurev.biophys.093008.131228} {\bibfield  {journal}
  {\bibinfo  {journal} {Annual Review of Biophysics}\ }\textbf {\bibinfo
  {volume} {39}},\ \bibinfo {pages} {265} (\bibinfo {year} {2010})}\BibitemShut
  {NoStop}%
\bibitem [{\citenamefont {Levine}\ and\ \citenamefont
  {Rappel}(2013)}]{Levine2013}%
  \BibitemOpen
  \bibfield  {author} {\bibinfo {author} {\bibfnamefont {H.}~\bibnamefont
  {Levine}}\ and\ \bibinfo {author} {\bibfnamefont {W.-J.}\ \bibnamefont
  {Rappel}},\ }\href@noop {} {\bibfield  {journal} {\bibinfo  {journal}
  {Physics Today}\ }\textbf {\bibinfo {volume} {66}},\ \bibinfo {pages} {24}
  (\bibinfo {year} {2013})}\BibitemShut {NoStop}%
\bibitem [{\citenamefont {Artemenko}\ \emph {et~al.}(2014)\citenamefont
  {Artemenko}, \citenamefont {Lampert},\ and\ \citenamefont
  {Devreotes}}]{Artemenko2014}%
  \BibitemOpen
  \bibfield  {author} {\bibinfo {author} {\bibfnamefont {Y.}~\bibnamefont
  {Artemenko}}, \bibinfo {author} {\bibfnamefont {T.~J.}\ \bibnamefont
  {Lampert}}, \ and\ \bibinfo {author} {\bibfnamefont {P.~N.}\ \bibnamefont
  {Devreotes}},\ }\href {\doibase 10.1007/s00018-014-1638-8} {\bibfield
  {journal} {\bibinfo  {journal} {Cellular and molecular life sciences : CMLS}\
  }\textbf {\bibinfo {volume} {71}},\ \bibinfo {pages} {3711} (\bibinfo {year}
  {2014})}\BibitemShut {NoStop}%
\bibitem [{\citenamefont {Roussos}\ \emph {et~al.}(2011)\citenamefont
  {Roussos}, \citenamefont {Condeelis},\ and\ \citenamefont
  {Patsialou}}]{Roussos2011}%
  \BibitemOpen
  \bibfield  {author} {\bibinfo {author} {\bibfnamefont {E.~T.}\ \bibnamefont
  {Roussos}}, \bibinfo {author} {\bibfnamefont {J.~S.}\ \bibnamefont
  {Condeelis}}, \ and\ \bibinfo {author} {\bibfnamefont {A.}~\bibnamefont
  {Patsialou}},\ }\href@noop {} {\bibfield  {journal} {\bibinfo  {journal} {Nat
  Rev Cancer}\ }\textbf {\bibinfo {volume} {11}},\ \bibinfo {pages} {573}
  (\bibinfo {year} {2011})}\BibitemShut {NoStop}%
\bibitem [{\citenamefont {Jarrell}\ and\ \citenamefont
  {McBride}(2008)}]{Jarrell2008}%
  \BibitemOpen
  \bibfield  {author} {\bibinfo {author} {\bibfnamefont {K.~F.}\ \bibnamefont
  {Jarrell}}\ and\ \bibinfo {author} {\bibfnamefont {M.~J.}\ \bibnamefont
  {McBride}},\ }\href {\doibase 10.1038/nrmicro1900} {\bibfield  {journal}
  {\bibinfo  {journal} {Nature Reviews Microbiology}\ }\textbf {\bibinfo
  {volume} {6}},\ \bibinfo {pages} {466} (\bibinfo {year} {2008})}\BibitemShut
  {NoStop}%
\bibitem [{\citenamefont {Wan}\ and\ \citenamefont {J\'ekely}(2021)}]{Wan2020}%
  \BibitemOpen
  \bibfield  {author} {\bibinfo {author} {\bibfnamefont {K.~Y.}\ \bibnamefont
  {Wan}}\ and\ \bibinfo {author} {\bibfnamefont {G.}~\bibnamefont {J\'ekely}},\
  }\href {\doibase 10.1098/rstb.2019.0758} {\bibfield  {journal} {\bibinfo
  {journal} {Philosophical Transactions of the Royal Society B: Biological
  Sciences}\ }\textbf {\bibinfo {volume} {376}},\ \bibinfo {pages} {20190758}
  (\bibinfo {year} {2021})},\ \Eprint
  {http://arxiv.org/abs/https://royalsocietypublishing.org/doi/pdf/10.1098/rstb.2019.0758}
  {https://royalsocietypublishing.org/doi/pdf/10.1098/rstb.2019.0758}
  \BibitemShut {NoStop}%
\bibitem [{\citenamefont {Butler}\ \emph {et~al.}(2018)\citenamefont {Butler},
  \citenamefont {Davies}, \citenamefont {Cartwright}, \citenamefont {Isayev},\
  and\ \citenamefont {Walsh}}]{Butler2018}%
  \BibitemOpen
  \bibfield  {author} {\bibinfo {author} {\bibfnamefont {K.~T.}\ \bibnamefont
  {Butler}}, \bibinfo {author} {\bibfnamefont {D.~W.}\ \bibnamefont {Davies}},
  \bibinfo {author} {\bibfnamefont {H.}~\bibnamefont {Cartwright}}, \bibinfo
  {author} {\bibfnamefont {O.}~\bibnamefont {Isayev}}, \ and\ \bibinfo {author}
  {\bibfnamefont {A.}~\bibnamefont {Walsh}},\ }\href {\doibase
  10.1038/s41586-018-0337-2} {\bibfield  {journal} {\bibinfo  {journal}
  {Nature}\ }\textbf {\bibinfo {volume} {559}},\ \bibinfo {pages} {547}
  (\bibinfo {year} {2018})}\BibitemShut {NoStop}%
\bibitem [{\citenamefont {Mehta}\ \emph {et~al.}(2019)\citenamefont {Mehta},
  \citenamefont {Bukov}, \citenamefont {Wang}, \citenamefont {Day},
  \citenamefont {Richardson}, \citenamefont {Fisher},\ and\ \citenamefont
  {Schwab}}]{Mehta2019}%
  \BibitemOpen
  \bibfield  {author} {\bibinfo {author} {\bibfnamefont {P.}~\bibnamefont
  {Mehta}}, \bibinfo {author} {\bibfnamefont {M.}~\bibnamefont {Bukov}},
  \bibinfo {author} {\bibfnamefont {C.~H.}\ \bibnamefont {Wang}}, \bibinfo
  {author} {\bibfnamefont {A.~G.}\ \bibnamefont {Day}}, \bibinfo {author}
  {\bibfnamefont {C.}~\bibnamefont {Richardson}}, \bibinfo {author}
  {\bibfnamefont {C.~K.}\ \bibnamefont {Fisher}}, \ and\ \bibinfo {author}
  {\bibfnamefont {D.~J.}\ \bibnamefont {Schwab}},\ }\href {\doibase
  10.1016/j.physrep.2019.03.001} {\bibfield  {journal} {\bibinfo  {journal}
  {Physics Reports}\ }\textbf {\bibinfo {volume} {810}},\ \bibinfo {pages} {1}
  (\bibinfo {year} {2019})},\ \Eprint {http://arxiv.org/abs/1803.08823}
  {arXiv:1803.08823} \BibitemShut {NoStop}%
\bibitem [{\citenamefont {Brunton}\ \emph {et~al.}(2020)\citenamefont
  {Brunton}, \citenamefont {Noack},\ and\ \citenamefont
  {Koumoutsakos}}]{Brunton2020}%
  \BibitemOpen
  \bibfield  {author} {\bibinfo {author} {\bibfnamefont {S.~L.}\ \bibnamefont
  {Brunton}}, \bibinfo {author} {\bibfnamefont {B.~R.}\ \bibnamefont {Noack}},
  \ and\ \bibinfo {author} {\bibfnamefont {P.}~\bibnamefont {Koumoutsakos}},\
  }\href {\doibase 10.1146/annurev-fluid-010719-060214} {\bibfield  {journal}
  {\bibinfo  {journal} {Annual Review of Fluid Mechanics}\ }\textbf {\bibinfo
  {volume} {52}},\ \bibinfo {pages} {477} (\bibinfo {year} {2020})},\ \Eprint
  {http://arxiv.org/abs/1905.11075} {arXiv:1905.11075} \BibitemShut {NoStop}%
\bibitem [{\citenamefont {Cichos}\ \emph {et~al.}(2020)\citenamefont {Cichos},
  \citenamefont {Gustavsson}, \citenamefont {Mehlig},\ and\ \citenamefont
  {Volpe}}]{Cichos2020}%
  \BibitemOpen
  \bibfield  {author} {\bibinfo {author} {\bibfnamefont {F.}~\bibnamefont
  {Cichos}}, \bibinfo {author} {\bibfnamefont {K.}~\bibnamefont {Gustavsson}},
  \bibinfo {author} {\bibfnamefont {B.}~\bibnamefont {Mehlig}}, \ and\ \bibinfo
  {author} {\bibfnamefont {G.}~\bibnamefont {Volpe}},\ }\href {\doibase
  10.1038/s42256-020-0146-9} {\bibfield  {journal} {\bibinfo  {journal} {Nature
  Machine Intelligence}\ }\textbf {\bibinfo {volume} {2}},\ \bibinfo {pages}
  {94} (\bibinfo {year} {2020})}\BibitemShut {NoStop}%
\bibitem [{\citenamefont {Colabrese}\ \emph {et~al.}(2017)\citenamefont
  {Colabrese}, \citenamefont {Gustavsson}, \citenamefont {Celani},\ and\
  \citenamefont {Biferale}}]{Colabrese2017}%
  \BibitemOpen
  \bibfield  {author} {\bibinfo {author} {\bibfnamefont {S.}~\bibnamefont
  {Colabrese}}, \bibinfo {author} {\bibfnamefont {K.}~\bibnamefont
  {Gustavsson}}, \bibinfo {author} {\bibfnamefont {A.}~\bibnamefont {Celani}},
  \ and\ \bibinfo {author} {\bibfnamefont {L.}~\bibnamefont {Biferale}},\
  }\href {\doibase 10.1103/PhysRevLett.118.158004} {\bibfield  {journal}
  {\bibinfo  {journal} {Physical Review Letters}\ }\textbf {\bibinfo {volume}
  {118}},\ \bibinfo {pages} {1} (\bibinfo {year} {2017})},\ \Eprint
  {http://arxiv.org/abs/1701.08848} {arXiv:1701.08848} \BibitemShut {NoStop}%
\bibitem [{\citenamefont {Gustavsson}\ \emph {et~al.}(2017)\citenamefont
  {Gustavsson}, \citenamefont {Biferale}, \citenamefont {Celani},\ and\
  \citenamefont {Colabrese}}]{Gustavsson2017}%
  \BibitemOpen
  \bibfield  {author} {\bibinfo {author} {\bibfnamefont {K.}~\bibnamefont
  {Gustavsson}}, \bibinfo {author} {\bibfnamefont {L.}~\bibnamefont
  {Biferale}}, \bibinfo {author} {\bibfnamefont {A.}~\bibnamefont {Celani}}, \
  and\ \bibinfo {author} {\bibfnamefont {S.}~\bibnamefont {Colabrese}},\ }\href
  {\doibase 10.1140/epje/i2017-11602-9} {\bibfield  {journal} {\bibinfo
  {journal} {European Physical Journal E}\ }\textbf {\bibinfo {volume} {40}}
  (\bibinfo {year} {2017}),\ 10.1140/epje/i2017-11602-9},\ \Eprint
  {http://arxiv.org/abs/1711.05826} {arXiv:1711.05826} \BibitemShut {NoStop}%
\bibitem [{\citenamefont {Alageshan}\ \emph {et~al.}(2020)\citenamefont
  {Alageshan}, \citenamefont {Verma}, \citenamefont {Bec},\ and\ \citenamefont
  {Pandit}}]{Alageshan2020}%
  \BibitemOpen
  \bibfield  {author} {\bibinfo {author} {\bibfnamefont {J.~K.}\ \bibnamefont
  {Alageshan}}, \bibinfo {author} {\bibfnamefont {A.~K.}\ \bibnamefont
  {Verma}}, \bibinfo {author} {\bibfnamefont {J.}~\bibnamefont {Bec}}, \ and\
  \bibinfo {author} {\bibfnamefont {R.}~\bibnamefont {Pandit}},\ }\href
  {\doibase 10.1103/PhysRevE.101.043110} {\bibfield  {journal} {\bibinfo
  {journal} {Physical Review E}\ }\textbf {\bibinfo {volume} {101}},\ \bibinfo
  {pages} {43110} (\bibinfo {year} {2020})}\BibitemShut {NoStop}%
\bibitem [{\citenamefont {Qiu}\ \emph {et~al.}(2020)\citenamefont {Qiu},
  \citenamefont {Huang}, \citenamefont {Xu},\ and\ \citenamefont
  {Zhao}}]{Qiu2020}%
  \BibitemOpen
  \bibfield  {author} {\bibinfo {author} {\bibfnamefont {J.~R.}\ \bibnamefont
  {Qiu}}, \bibinfo {author} {\bibfnamefont {W.~X.}\ \bibnamefont {Huang}},
  \bibinfo {author} {\bibfnamefont {C.~X.}\ \bibnamefont {Xu}}, \ and\ \bibinfo
  {author} {\bibfnamefont {L.~H.}\ \bibnamefont {Zhao}},\ }\href {\doibase
  10.1007/s11433-019-1502-2} {\bibfield  {journal} {\bibinfo  {journal}
  {Science China: Physics, Mechanics and Astronomy}\ }\textbf {\bibinfo
  {volume} {63}} (\bibinfo {year} {2020}),\ 10.1007/s11433-019-1502-2},\
  \Eprint {http://arxiv.org/abs/1811.10880} {arXiv:1811.10880} \BibitemShut
  {NoStop}%
\bibitem [{\citenamefont {Reddy}\ \emph {et~al.}(2016)\citenamefont {Reddy},
  \citenamefont {Celani}, \citenamefont {Sejnowski},\ and\ \citenamefont
  {Vergassola}}]{Reddy2016}%
  \BibitemOpen
  \bibfield  {author} {\bibinfo {author} {\bibfnamefont {G.}~\bibnamefont
  {Reddy}}, \bibinfo {author} {\bibfnamefont {A.}~\bibnamefont {Celani}},
  \bibinfo {author} {\bibfnamefont {T.~J.}\ \bibnamefont {Sejnowski}}, \ and\
  \bibinfo {author} {\bibfnamefont {M.}~\bibnamefont {Vergassola}},\ }\href
  {\doibase 10.1073/pnas.1606075113} {\bibfield  {journal} {\bibinfo  {journal}
  {Proceedings of the National Academy of Sciences of the United States of
  America}\ }\textbf {\bibinfo {volume} {113}},\ \bibinfo {pages} {E4877}
  (\bibinfo {year} {2016})}\BibitemShut {NoStop}%
\bibitem [{\citenamefont {Palmer}\ and\ \citenamefont
  {Yaida}(2017)}]{Palmer2017}%
  \BibitemOpen
  \bibfield  {author} {\bibinfo {author} {\bibfnamefont {G.}~\bibnamefont
  {Palmer}}\ and\ \bibinfo {author} {\bibfnamefont {S.}~\bibnamefont {Yaida}},\
  }\href {http://arxiv.org/abs/1709.02379} {\bibfield  {journal} {\bibinfo
  {journal} {Arxiv preprint}\ ,\ \bibinfo {pages} {arXiv:1709.02379}} (\bibinfo
  {year} {2017})},\ \Eprint {http://arxiv.org/abs/1709.02379}
  {arXiv:1709.02379} \BibitemShut {NoStop}%
\bibitem [{\citenamefont {Schneider}\ and\ \citenamefont
  {Stark}(2019)}]{Schneider2019}%
  \BibitemOpen
  \bibfield  {author} {\bibinfo {author} {\bibfnamefont {E.}~\bibnamefont
  {Schneider}}\ and\ \bibinfo {author} {\bibfnamefont {H.}~\bibnamefont
  {Stark}},\ }\href {\doibase 10.1209/0295-5075/127/64003} {\bibfield
  {journal} {\bibinfo  {journal} {Epl}\ }\textbf {\bibinfo {volume} {127}}
  (\bibinfo {year} {2019}),\ 10.1209/0295-5075/127/64003},\ \Eprint
  {http://arxiv.org/abs/1909.03243} {arXiv:1909.03243} \BibitemShut {NoStop}%
\bibitem [{\citenamefont {Tsang}\ \emph {et~al.}(2020)\citenamefont {Tsang},
  \citenamefont {Tong}, \citenamefont {Nallan},\ and\ \citenamefont
  {Pak}}]{Tsang2018}%
  \BibitemOpen
  \bibfield  {author} {\bibinfo {author} {\bibfnamefont {A.~C.~H.}\
  \bibnamefont {Tsang}}, \bibinfo {author} {\bibfnamefont {P.~W.}\ \bibnamefont
  {Tong}}, \bibinfo {author} {\bibfnamefont {S.}~\bibnamefont {Nallan}}, \ and\
  \bibinfo {author} {\bibfnamefont {O.~S.}\ \bibnamefont {Pak}},\ }\href
  {https://doi.org/10.1103/PhysRevFluids.5.074101} {\bibfield  {journal}
  {\bibinfo  {journal} {Phys. Rev. Fluids}\ }\textbf {\bibinfo {volume} {5}},\
  \bibinfo {pages} {074101} (\bibinfo {year} {2020})},\ \Eprint
  {http://arxiv.org/abs/1808.07639} {arXiv:1808.07639} \BibitemShut {NoStop}%
\bibitem [{\citenamefont {Verma}\ \emph {et~al.}(2018)\citenamefont {Verma},
  \citenamefont {Novati},\ and\ \citenamefont {Koumoutsakos}}]{Verma2018}%
  \BibitemOpen
  \bibfield  {author} {\bibinfo {author} {\bibfnamefont {S.}~\bibnamefont
  {Verma}}, \bibinfo {author} {\bibfnamefont {G.}~\bibnamefont {Novati}}, \
  and\ \bibinfo {author} {\bibfnamefont {P.}~\bibnamefont {Koumoutsakos}},\
  }\href {\doibase 10.1073/pnas.1800923115} {\bibfield  {journal} {\bibinfo
  {journal} {Proceedings of the National Academy of Sciences of the United
  States of America}\ }\textbf {\bibinfo {volume} {115}},\ \bibinfo {pages}
  {5849} (\bibinfo {year} {2018})},\ \Eprint {http://arxiv.org/abs/1802.02674}
  {arXiv:1802.02674} \BibitemShut {NoStop}%
\bibitem [{\citenamefont {Mui{\~{n}}os-Landin}\ \emph
  {et~al.}(2018)\citenamefont {Mui{\~{n}}os-Landin}, \citenamefont
  {Ghazi-Zahedi},\ and\ \citenamefont {Cichos}}]{Muinos-Landin2018}%
  \BibitemOpen
  \bibfield  {author} {\bibinfo {author} {\bibfnamefont {S.}~\bibnamefont
  {Mui{\~{n}}os-Landin}}, \bibinfo {author} {\bibfnamefont {K.}~\bibnamefont
  {Ghazi-Zahedi}}, \ and\ \bibinfo {author} {\bibfnamefont {F.}~\bibnamefont
  {Cichos}},\ }\href {http://arxiv.org/abs/1803.06425} {\bibfield  {journal}
  {\bibinfo  {journal} {Arxiv preprint}\ ,\ \bibinfo {pages}
  {arXiv:1803.06425}} (\bibinfo {year} {2018})},\ \Eprint
  {http://arxiv.org/abs/1803.06425} {arXiv:1803.06425} \BibitemShut {NoStop}%
\bibitem [{\citenamefont {Reddy}\ \emph {et~al.}(2018)\citenamefont {Reddy},
  \citenamefont {Wong-Ng}, \citenamefont {Celani}, \citenamefont {Sejnowski},\
  and\ \citenamefont {Vergassola}}]{Reddy2018}%
  \BibitemOpen
  \bibfield  {author} {\bibinfo {author} {\bibfnamefont {G.}~\bibnamefont
  {Reddy}}, \bibinfo {author} {\bibfnamefont {J.}~\bibnamefont {Wong-Ng}},
  \bibinfo {author} {\bibfnamefont {A.}~\bibnamefont {Celani}}, \bibinfo
  {author} {\bibfnamefont {T.~J.}\ \bibnamefont {Sejnowski}}, \ and\ \bibinfo
  {author} {\bibfnamefont {M.}~\bibnamefont {Vergassola}},\ }\href {\doibase
  10.1038/s41586-018-0533-0} {\bibfield  {journal} {\bibinfo  {journal}
  {Nature}\ }\textbf {\bibinfo {volume} {562}},\ \bibinfo {pages} {236}
  (\bibinfo {year} {2018})}\BibitemShut {NoStop}%
\bibitem [{\citenamefont {Najafi}\ and\ \citenamefont
  {Golestanian}(2004)}]{Najafi2004}%
  \BibitemOpen
  \bibfield  {author} {\bibinfo {author} {\bibfnamefont {A.}~\bibnamefont
  {Najafi}}\ and\ \bibinfo {author} {\bibfnamefont {R.}~\bibnamefont
  {Golestanian}},\ }\href {\doibase 10.1103/PhysRevE.69.062901} {\bibfield
  {journal} {\bibinfo  {journal} {Phys. Rev. E}\ }\textbf {\bibinfo {volume}
  {69}},\ \bibinfo {pages} {062901} (\bibinfo {year} {2004})},\ \Eprint
  {http://arxiv.org/abs/0402070} {arXiv:0402070 [cond-mat]} \BibitemShut
  {NoStop}%
\bibitem [{\citenamefont {Golestanian}\ and\ \citenamefont
  {Ajdari}(2008)}]{Golestanian2008}%
  \BibitemOpen
  \bibfield  {author} {\bibinfo {author} {\bibfnamefont {R.}~\bibnamefont
  {Golestanian}}\ and\ \bibinfo {author} {\bibfnamefont {A.}~\bibnamefont
  {Ajdari}},\ }\href {\doibase 10.1103/PhysRevE.77.036308} {\bibfield
  {journal} {\bibinfo  {journal} {Physical Review E - Statistical, Nonlinear,
  and Soft Matter Physics}\ }\textbf {\bibinfo {volume} {77}},\ \bibinfo
  {pages} {1} (\bibinfo {year} {2008})},\ \Eprint
  {http://arxiv.org/abs/0711.3700} {arXiv:0711.3700} \BibitemShut {NoStop}%
\bibitem [{\citenamefont {Earl}\ \emph {et~al.}(2007)\citenamefont {Earl},
  \citenamefont {Pooley}, \citenamefont {Ryder}, \citenamefont {Bredberg},\
  and\ \citenamefont {Yeomans}}]{Earl2007}%
  \BibitemOpen
  \bibfield  {author} {\bibinfo {author} {\bibfnamefont {D.~J.}\ \bibnamefont
  {Earl}}, \bibinfo {author} {\bibfnamefont {C.~M.}\ \bibnamefont {Pooley}},
  \bibinfo {author} {\bibfnamefont {J.~F.}\ \bibnamefont {Ryder}}, \bibinfo
  {author} {\bibfnamefont {I.}~\bibnamefont {Bredberg}}, \ and\ \bibinfo
  {author} {\bibfnamefont {J.~M.}\ \bibnamefont {Yeomans}},\ }\href {\doibase
  10.1063/1.2434160} {\bibfield  {journal} {\bibinfo  {journal} {J. Chem.
  Phys.}\ }\textbf {\bibinfo {volume} {126}},\ \bibinfo {pages} {064703}
  (\bibinfo {year} {2007})},\ \Eprint {http://arxiv.org/abs/0701511}
  {arXiv:0701511 [cond-mat]} \BibitemShut {NoStop}%
\bibitem [{\citenamefont {Sutton}\ and\ \citenamefont
  {Barto}(2018)}]{Sutton1998}%
  \BibitemOpen
  \bibfield  {author} {\bibinfo {author} {\bibfnamefont {R.~S.}\ \bibnamefont
  {Sutton}}\ and\ \bibinfo {author} {\bibfnamefont {A.~G.}\ \bibnamefont
  {Barto}},\ }\href {http://incompleteideas.net/book/the-book-2nd.html} {\emph
  {\bibinfo {title} {{Reinforcement Learning: An Introduction}}}},\ \bibinfo
  {edition} {2nd}\ ed.\ (\bibinfo  {publisher} {The MIT Press},\ \bibinfo
  {year} {2018})\BibitemShut {NoStop}%
\bibitem [{\citenamefont {Mnih}\ \emph {et~al.}(2013)\citenamefont {Mnih},
  \citenamefont {Kavukcuoglu}, \citenamefont {Silver}, \citenamefont {Graves},
  \citenamefont {Antonoglou}, \citenamefont {Wierstra},\ and\ \citenamefont
  {Riedmiller}}]{Mnih2013}%
  \BibitemOpen
  \bibfield  {author} {\bibinfo {author} {\bibfnamefont {V.}~\bibnamefont
  {Mnih}}, \bibinfo {author} {\bibfnamefont {K.}~\bibnamefont {Kavukcuoglu}},
  \bibinfo {author} {\bibfnamefont {D.}~\bibnamefont {Silver}}, \bibinfo
  {author} {\bibfnamefont {A.}~\bibnamefont {Graves}}, \bibinfo {author}
  {\bibfnamefont {I.}~\bibnamefont {Antonoglou}}, \bibinfo {author}
  {\bibfnamefont {D.}~\bibnamefont {Wierstra}}, \ and\ \bibinfo {author}
  {\bibfnamefont {M.}~\bibnamefont {Riedmiller}},\ }\href
  {http://arxiv.org/abs/1312.5602} {\bibfield  {journal} {\bibinfo  {journal}
  {arxiv}\ } (\bibinfo {year} {2013})}\BibitemShut {NoStop}%
\bibitem [{\citenamefont {Senior}\ \emph {et~al.}(2020)\citenamefont {Senior},
  \citenamefont {Evans}, \citenamefont {Jumper}, \citenamefont {Kirkpatrick},
  \citenamefont {Sifre}, \citenamefont {Green}, \citenamefont {Qin},
  \citenamefont {{\v{Z}}{\'{i}}dek}, \citenamefont {Nelson}, \citenamefont
  {Bridgland}, \citenamefont {Penedones}, \citenamefont {Petersen},
  \citenamefont {Simonyan}, \citenamefont {Crossan}, \citenamefont {Kohli},
  \citenamefont {Jones}, \citenamefont {Silver}, \citenamefont {Kavukcuoglu},\
  and\ \citenamefont {Hassabis}}]{Senior2020}%
  \BibitemOpen
  \bibfield  {author} {\bibinfo {author} {\bibfnamefont {A.~W.}\ \bibnamefont
  {Senior}}, \bibinfo {author} {\bibfnamefont {R.}~\bibnamefont {Evans}},
  \bibinfo {author} {\bibfnamefont {J.}~\bibnamefont {Jumper}}, \bibinfo
  {author} {\bibfnamefont {J.}~\bibnamefont {Kirkpatrick}}, \bibinfo {author}
  {\bibfnamefont {L.}~\bibnamefont {Sifre}}, \bibinfo {author} {\bibfnamefont
  {T.}~\bibnamefont {Green}}, \bibinfo {author} {\bibfnamefont
  {C.}~\bibnamefont {Qin}}, \bibinfo {author} {\bibfnamefont {A.}~\bibnamefont
  {{\v{Z}}{\'{i}}dek}}, \bibinfo {author} {\bibfnamefont {A.~W.~R.}\
  \bibnamefont {Nelson}}, \bibinfo {author} {\bibfnamefont {A.}~\bibnamefont
  {Bridgland}}, \bibinfo {author} {\bibfnamefont {H.}~\bibnamefont
  {Penedones}}, \bibinfo {author} {\bibfnamefont {S.}~\bibnamefont {Petersen}},
  \bibinfo {author} {\bibfnamefont {K.}~\bibnamefont {Simonyan}}, \bibinfo
  {author} {\bibfnamefont {S.}~\bibnamefont {Crossan}}, \bibinfo {author}
  {\bibfnamefont {P.}~\bibnamefont {Kohli}}, \bibinfo {author} {\bibfnamefont
  {D.~T.}\ \bibnamefont {Jones}}, \bibinfo {author} {\bibfnamefont
  {D.}~\bibnamefont {Silver}}, \bibinfo {author} {\bibfnamefont
  {K.}~\bibnamefont {Kavukcuoglu}}, \ and\ \bibinfo {author} {\bibfnamefont
  {D.}~\bibnamefont {Hassabis}},\ }\href {\doibase 10.1038/s41586-019-1923-7}
  {\bibfield  {journal} {\bibinfo  {journal} {Nature}\ }\textbf {\bibinfo
  {volume} {577}},\ \bibinfo {pages} {706} (\bibinfo {year}
  {2020})}\BibitemShut {NoStop}%
\bibitem [{\citenamefont {Hochreiter}\ and\ \citenamefont
  {Schmidhuber}(1997)}]{Hochreiter1997}%
  \BibitemOpen
  \bibfield  {author} {\bibinfo {author} {\bibfnamefont {S.}~\bibnamefont
  {Hochreiter}}\ and\ \bibinfo {author} {\bibfnamefont {J.}~\bibnamefont
  {Schmidhuber}},\ }\href {\doibase 10.1162/neco.1997.9.8.1735} {\bibfield
  {journal} {\bibinfo  {journal} {Neural Comput.}\ }\textbf {\bibinfo {volume}
  {9}},\ \bibinfo {pages} {1735} (\bibinfo {year} {1997})}\BibitemShut
  {NoStop}%
\bibitem [{\citenamefont {Staudemeyer}\ and\ \citenamefont
  {Morris}(2019)}]{Staudmeyer2019}%
  \BibitemOpen
  \bibfield  {author} {\bibinfo {author} {\bibfnamefont {R.~C.}\ \bibnamefont
  {Staudemeyer}}\ and\ \bibinfo {author} {\bibfnamefont {E.~R.}\ \bibnamefont
  {Morris}},\ }\href {http://arxiv.org/abs/1909.09586} {\bibfield  {journal}
  {\bibinfo  {journal} {Arxiv preprint}\ ,\ \bibinfo {pages}
  {arXiv:1909.09586}} (\bibinfo {year} {2019})},\ \Eprint
  {http://arxiv.org/abs/1909.09586} {arXiv:1909.09586} \BibitemShut {NoStop}%
\bibitem [{\citenamefont {Theves}\ \emph {et~al.}(2013)\citenamefont {Theves},
  \citenamefont {Taktikos}, \citenamefont {Zaburdaev}, \citenamefont {Stark},\
  and\ \citenamefont {Beta}}]{Theves2013}%
  \BibitemOpen
  \bibfield  {author} {\bibinfo {author} {\bibfnamefont {M.}~\bibnamefont
  {Theves}}, \bibinfo {author} {\bibfnamefont {J.}~\bibnamefont {Taktikos}},
  \bibinfo {author} {\bibfnamefont {V.}~\bibnamefont {Zaburdaev}}, \bibinfo
  {author} {\bibfnamefont {H.}~\bibnamefont {Stark}}, \ and\ \bibinfo {author}
  {\bibfnamefont {C.}~\bibnamefont {Beta}},\ }\href {\doibase
  10.1016/j.bpj.2013.08.047} {\bibfield  {journal} {\bibinfo  {journal}
  {Biophysical Journal}\ }\textbf {\bibinfo {volume} {105}},\ \bibinfo {pages}
  {1915} (\bibinfo {year} {2013})}\BibitemShut {NoStop}%
\bibitem [{\citenamefont {Lillicrap}\ \emph {et~al.}(2015)\citenamefont
  {Lillicrap}, \citenamefont {Hunt}, \citenamefont {Pritzel}, \citenamefont
  {Heess}, \citenamefont {Erez}, \citenamefont {Tassa}, \citenamefont
  {Silver},\ and\ \citenamefont {Wierstra}}]{Lillicrap2015}%
  \BibitemOpen
  \bibfield  {author} {\bibinfo {author} {\bibfnamefont {T.~P.}\ \bibnamefont
  {Lillicrap}}, \bibinfo {author} {\bibfnamefont {J.~J.}\ \bibnamefont {Hunt}},
  \bibinfo {author} {\bibfnamefont {A.}~\bibnamefont {Pritzel}}, \bibinfo
  {author} {\bibfnamefont {N.}~\bibnamefont {Heess}}, \bibinfo {author}
  {\bibfnamefont {T.}~\bibnamefont {Erez}}, \bibinfo {author} {\bibfnamefont
  {Y.}~\bibnamefont {Tassa}}, \bibinfo {author} {\bibfnamefont
  {D.}~\bibnamefont {Silver}}, \ and\ \bibinfo {author} {\bibfnamefont
  {D.}~\bibnamefont {Wierstra}},\ }\href@noop {} {\bibfield  {journal}
  {\bibinfo  {journal} {Arxiv preprint}\ ,\ \bibinfo {pages}
  {arXiv:1509.02971}} (\bibinfo {year} {2015})},\ \Eprint
  {http://arxiv.org/abs/1509.02971} {arXiv:1509.02971 [cs.LG]} \BibitemShut
  {NoStop}%
\bibitem [{\citenamefont {Gu}\ \emph {et~al.}(2019)\citenamefont {Gu},
  \citenamefont {Jia},\ and\ \citenamefont {Choset}}]{Gu2019}%
  \BibitemOpen
  \bibfield  {author} {\bibinfo {author} {\bibfnamefont {Z.}~\bibnamefont
  {Gu}}, \bibinfo {author} {\bibfnamefont {Z.}~\bibnamefont {Jia}}, \ and\
  \bibinfo {author} {\bibfnamefont {H.}~\bibnamefont {Choset}},\ }\href@noop {}
  {\enquote {\bibinfo {title} {{Adversary A3C for Robust Reinforcement
  Learning}},}\ } (\bibinfo {year} {2019}),\ \Eprint
  {http://arxiv.org/abs/1912.00330} {arXiv:1912.00330 [cs.LG]} \BibitemShut
  {NoStop}%
\bibitem [{\citenamefont {Schulman}\ \emph {et~al.}(2017)\citenamefont
  {Schulman}, \citenamefont {Wolski}, \citenamefont {Dhariwal}, \citenamefont
  {Radford},\ and\ \citenamefont {Klimov}}]{Schulman2017}%
  \BibitemOpen
  \bibfield  {author} {\bibinfo {author} {\bibfnamefont {J.}~\bibnamefont
  {Schulman}}, \bibinfo {author} {\bibfnamefont {F.}~\bibnamefont {Wolski}},
  \bibinfo {author} {\bibfnamefont {P.}~\bibnamefont {Dhariwal}}, \bibinfo
  {author} {\bibfnamefont {A.}~\bibnamefont {Radford}}, \ and\ \bibinfo
  {author} {\bibfnamefont {O.}~\bibnamefont {Klimov}},\ }\href@noop {}
  {\enquote {\bibinfo {title} {{Proximal Policy Optimization Algorithms}},}\ }
  (\bibinfo {year} {2017}),\ \Eprint {http://arxiv.org/abs/1707.06347}
  {arXiv:1707.06347 [cs.LG]} \BibitemShut {NoStop}%
\bibitem [{\citenamefont {Watkins}\ and\ \citenamefont
  {Dayan}(1992)}]{Watkins1992}%
  \BibitemOpen
  \bibfield  {author} {\bibinfo {author} {\bibfnamefont {C.~J. C.~H.}\
  \bibnamefont {Watkins}}\ and\ \bibinfo {author} {\bibfnamefont
  {P.}~\bibnamefont {Dayan}},\ }\href {\doibase 10.1007/BF00992698} {\bibfield
  {journal} {\bibinfo  {journal} {Mach. Learn.}\ }\textbf {\bibinfo {volume}
  {8}},\ \bibinfo {pages} {279} (\bibinfo {year} {1992})}\BibitemShut {NoStop}%
\bibitem [{\citenamefont {Berg}\ and\ \citenamefont
  {Purcell}(1977)}]{Berg1977}%
  \BibitemOpen
  \bibfield  {author} {\bibinfo {author} {\bibfnamefont {H.~C.}\ \bibnamefont
  {Berg}}\ and\ \bibinfo {author} {\bibfnamefont {E.~M.}\ \bibnamefont
  {Purcell}},\ }\href {\doibase 10.1016/S0006-3495(77)85544-6} {\bibfield
  {journal} {\bibinfo  {journal} {Biophys. J.}\ }\textbf {\bibinfo {volume}
  {20}},\ \bibinfo {pages} {193} (\bibinfo {year} {1977})}\BibitemShut
  {NoStop}%
\bibitem [{\citenamefont {ten Wolde}\ \emph {et~al.}(2016)\citenamefont {ten
  Wolde}, \citenamefont {Becker}, \citenamefont {Ouldridge},\ and\
  \citenamefont {Mugler}}]{TenWolde2016a}%
  \BibitemOpen
  \bibfield  {author} {\bibinfo {author} {\bibfnamefont {P.~R.}\ \bibnamefont
  {ten Wolde}}, \bibinfo {author} {\bibfnamefont {N.~B.}\ \bibnamefont
  {Becker}}, \bibinfo {author} {\bibfnamefont {T.~E.}\ \bibnamefont
  {Ouldridge}}, \ and\ \bibinfo {author} {\bibfnamefont {A.}~\bibnamefont
  {Mugler}},\ }\href {\doibase 10.1007/s10955-015-1440-5} {\bibfield  {journal}
  {\bibinfo  {journal} {Journal of Statistical Physics}\ }\textbf {\bibinfo
  {volume} {162}},\ \bibinfo {pages} {1395} (\bibinfo {year} {2016})},\ \Eprint
  {http://arxiv.org/abs/1505.06577} {arXiv:1505.06577} \BibitemShut {NoStop}%
\bibitem [{\citenamefont {Bal{\'{a}}zsi}\ \emph {et~al.}(2011)\citenamefont
  {Bal{\'{a}}zsi}, \citenamefont {{Van Oudenaarden}},\ and\ \citenamefont
  {Collins}}]{Balazsi2011}%
  \BibitemOpen
  \bibfield  {author} {\bibinfo {author} {\bibfnamefont {G.}~\bibnamefont
  {Bal{\'{a}}zsi}}, \bibinfo {author} {\bibfnamefont {A.}~\bibnamefont {{Van
  Oudenaarden}}}, \ and\ \bibinfo {author} {\bibfnamefont {J.~J.}\ \bibnamefont
  {Collins}},\ }\href {\doibase 10.1016/j.cell.2011.01.030} {\bibfield
  {journal} {\bibinfo  {journal} {Cell}\ }\textbf {\bibinfo {volume} {144}},\
  \bibinfo {pages} {910} (\bibinfo {year} {2011})}\BibitemShut {NoStop}%
\bibitem [{\citenamefont {Bowsher}\ and\ \citenamefont
  {Swain}(2014)}]{Bowsher2014}%
  \BibitemOpen
  \bibfield  {author} {\bibinfo {author} {\bibfnamefont {C.~G.}\ \bibnamefont
  {Bowsher}}\ and\ \bibinfo {author} {\bibfnamefont {P.~S.}\ \bibnamefont
  {Swain}},\ }\href {\doibase 10.1016/j.copbio.2014.04.010} {\bibfield
  {journal} {\bibinfo  {journal} {Current Opinion in Biotechnology}\ }\textbf
  {\bibinfo {volume} {28}},\ \bibinfo {pages} {149} (\bibinfo {year}
  {2014})}\BibitemShut {NoStop}%
\bibitem [{\citenamefont {Tang}\ and\ \citenamefont
  {Marshall}(2018)}]{Tang2018}%
  \BibitemOpen
  \bibfield  {author} {\bibinfo {author} {\bibfnamefont {S.~K.}\ \bibnamefont
  {Tang}}\ and\ \bibinfo {author} {\bibfnamefont {W.~F.}\ \bibnamefont
  {Marshall}},\ }\href {\doibase 10.1016/j.cub.2018.09.015} {\bibfield
  {journal} {\bibinfo  {journal} {Current Biology}\ }\textbf {\bibinfo {volume}
  {28}},\ \bibinfo {pages} {R1180} (\bibinfo {year} {2018})}\BibitemShut
  {NoStop}%
\bibitem [{\citenamefont {Tripathi}\ \emph {et~al.}(2020)\citenamefont
  {Tripathi}, \citenamefont {Levine},\ and\ \citenamefont
  {Jolly}}]{Tripathi2020}%
  \BibitemOpen
  \bibfield  {author} {\bibinfo {author} {\bibfnamefont {S.}~\bibnamefont
  {Tripathi}}, \bibinfo {author} {\bibfnamefont {H.}~\bibnamefont {Levine}}, \
  and\ \bibinfo {author} {\bibfnamefont {M.~K.}\ \bibnamefont {Jolly}},\ }\href
  {\doibase 10.1146/annurev-biophys-121219-081557} {\bibfield  {journal}
  {\bibinfo  {journal} {Annual Review of Biophysics}\ }\textbf {\bibinfo
  {volume} {49}},\ \bibinfo {pages} {1} (\bibinfo {year} {2020})}\BibitemShut
  {NoStop}%
\bibitem [{\citenamefont {Reid}\ \emph {et~al.}(2015)\citenamefont {Reid},
  \citenamefont {Garnier}, \citenamefont {Beekman},\ and\ \citenamefont
  {Latty}}]{Reid2015}%
  \BibitemOpen
  \bibfield  {author} {\bibinfo {author} {\bibfnamefont {C.~R.}\ \bibnamefont
  {Reid}}, \bibinfo {author} {\bibfnamefont {S.}~\bibnamefont {Garnier}},
  \bibinfo {author} {\bibfnamefont {M.}~\bibnamefont {Beekman}}, \ and\
  \bibinfo {author} {\bibfnamefont {T.}~\bibnamefont {Latty}},\ }\href
  {\doibase 10.1016/j.anbehav.2014.11.010} {\bibfield  {journal} {\bibinfo
  {journal} {Animal Behaviour}\ }\textbf {\bibinfo {volume} {100}},\ \bibinfo
  {pages} {44} (\bibinfo {year} {2015})}\BibitemShut {NoStop}%
\bibitem [{\citenamefont {Jarrell}\ \emph {et~al.}(2012)\citenamefont
  {Jarrell}, \citenamefont {Wang}, \citenamefont {Bloniarz}, \citenamefont
  {Brittin}, \citenamefont {Xu}, \citenamefont {Thomson}, \citenamefont
  {Albertson}, \citenamefont {Hall},\ and\ \citenamefont
  {Emmons}}]{Jarrell2012}%
  \BibitemOpen
  \bibfield  {author} {\bibinfo {author} {\bibfnamefont {T.~A.}\ \bibnamefont
  {Jarrell}}, \bibinfo {author} {\bibfnamefont {Y.}~\bibnamefont {Wang}},
  \bibinfo {author} {\bibfnamefont {A.~E.}\ \bibnamefont {Bloniarz}}, \bibinfo
  {author} {\bibfnamefont {C.~A.}\ \bibnamefont {Brittin}}, \bibinfo {author}
  {\bibfnamefont {M.}~\bibnamefont {Xu}}, \bibinfo {author} {\bibfnamefont
  {J.~N.}\ \bibnamefont {Thomson}}, \bibinfo {author} {\bibfnamefont {D.~G.}\
  \bibnamefont {Albertson}}, \bibinfo {author} {\bibfnamefont {D.~H.}\
  \bibnamefont {Hall}}, \ and\ \bibinfo {author} {\bibfnamefont {S.~W.}\
  \bibnamefont {Emmons}},\ }\href {\doibase 10.1126/science.1221762} {\bibfield
   {journal} {\bibinfo  {journal} {Science}\ }\textbf {\bibinfo {volume}
  {337}},\ \bibinfo {pages} {437} (\bibinfo {year} {2012})}\BibitemShut
  {NoStop}%
\bibitem [{\citenamefont {Itskovits}\ \emph {et~al.}(2018)\citenamefont
  {Itskovits}, \citenamefont {Ruach},\ and\ \citenamefont
  {Zaslaver}}]{Itskovits2018}%
  \BibitemOpen
  \bibfield  {author} {\bibinfo {author} {\bibfnamefont {E.}~\bibnamefont
  {Itskovits}}, \bibinfo {author} {\bibfnamefont {R.}~\bibnamefont {Ruach}}, \
  and\ \bibinfo {author} {\bibfnamefont {A.}~\bibnamefont {Zaslaver}},\ }\href
  {\doibase 10.1038/s41467-018-05151-2} {\bibfield  {journal} {\bibinfo
  {journal} {Nature Communications}\ }\textbf {\bibinfo {volume} {9}} (\bibinfo
  {year} {2018}),\ 10.1038/s41467-018-05151-2}\BibitemShut {NoStop}%
\bibitem [{\citenamefont {Goodfellow}\ \emph {et~al.}(2016)\citenamefont
  {Goodfellow}, \citenamefont {Bengio},\ and\ \citenamefont
  {Courville}}]{Goodfellow2016}%
  \BibitemOpen
  \bibfield  {author} {\bibinfo {author} {\bibfnamefont {I.}~\bibnamefont
  {Goodfellow}}, \bibinfo {author} {\bibfnamefont {Y.}~\bibnamefont {Bengio}},
  \ and\ \bibinfo {author} {\bibfnamefont {A.}~\bibnamefont {Courville}},\
  }\href@noop {} {\emph {\bibinfo {title} {{Deep Learning}}}}\ (\bibinfo
  {publisher} {MIT Press},\ \bibinfo {year} {2016})\BibitemShut {NoStop}%
\bibitem [{\citenamefont {Baker}\ and\ \citenamefont
  {Patil}(1998)}]{Baker1998}%
  \BibitemOpen
  \bibfield  {author} {\bibinfo {author} {\bibfnamefont {M.~R.}\ \bibnamefont
  {Baker}}\ and\ \bibinfo {author} {\bibfnamefont {R.~B.}\ \bibnamefont
  {Patil}},\ }\href {\doibase 10.1023/A:1009951412412} {\bibfield  {journal}
  {\bibinfo  {journal} {Reliab. Comput.}\ }\textbf {\bibinfo {volume} {4}},\
  \bibinfo {pages} {235} (\bibinfo {year} {1998})}\BibitemShut {NoStop}%
\bibitem [{\citenamefont {Stanley}\ and\ \citenamefont
  {Miikkulainen}(2002)}]{Stanley2002}%
  \BibitemOpen
  \bibfield  {author} {\bibinfo {author} {\bibfnamefont {K.~O.}\ \bibnamefont
  {Stanley}}\ and\ \bibinfo {author} {\bibfnamefont {R.}~\bibnamefont
  {Miikkulainen}},\ }\href {\doibase 10.1162/106365602320169811} {\bibfield
  {journal} {\bibinfo  {journal} {Evol. Comput.}\ }\textbf {\bibinfo {volume}
  {10}},\ \bibinfo {pages} {99} (\bibinfo {year} {2002})}\BibitemShut {NoStop}%
\end{thebibliography}
%

\end{document}